\documentclass[fleqn,usenatbib]{mnras}

\usepackage{newtxtext,newtxmath}

\usepackage[T1]{fontenc}
\usepackage{ae,aecompl}
\usepackage{breqn}
\usepackage{mathtools}
\usepackage[utf8]{inputenc} 
\usepackage{gensymb}
\usepackage{upgreek}

\usepackage{graphicx}	
\usepackage{amsmath}	
\usepackage{amssymb}	
\usepackage{latexsym}
\usepackage{longtable}

\title[Radio observations of planetary nebulae]{Radio observations of planetary nebulae: no evidence for strong radial density gradients}

\author[M. Hajduk et al.]{M. Hajduk,$^{1}$\thanks{E-mail: marcin.hajduk@uwm.edu.pl}
P. A. M. van Hoof,$^{2}$
K. \'Sniadkowska,$^{1}$
A. Krankowski,$^{1}$ 
\newauthor L. B\l{}aszkiewicz,$^{1,3}$
B. D\k{a}browski,$^{1}$
and A. A. Zijlstra$^{4}$
\\
$^{1}$Space Radio-Diagnostics Research Centre, University of Warmia and Mazury in Olsztyn, Prawoche\'nskiego 9, 10-720 Olsztyn, Poland\\
$^{2}$Royal Observatory of Belgium, Ringlaan 3, B-1180 Brussels, Belgium\\
$^{3}$Faculty of Mathematics and Computer Sciences, University of Warmia and Mazury in Olsztyn, 
S\l{}oneczna 54, 10-720 Olsztyn, Poland\\
$^{4}$Department of Astronomy and Astrophysics, The University of Manchester, Manchester, M13 9PL, UK \\
}

\date{Accepted XXX. Received YYY; in original form ZZZ}

\pubyear{2018}

\begin{document}
\label{firstpage}
\pagerange{\pageref{firstpage}--\pageref{lastpage}}
\maketitle

\begin{abstract}
Radio continuum observations trace thermal emission of ionized plasma in planetary nebulae and bring useful information on nebular geometries. A model of homogeneous sphere or shell cannot fit the nebular spectra and brightness temperatures. Two alternative models have been proposed in the literature: the first one consists of two homogeneous components, while the other one is a model of a shell with a significant radial density gradient. On the other side, prolate ellipsoidal shell model can successfully fit the surface brightness distribution of selected objects. We verify the existing models using data collected in radio surveys covering wide range of frequencies. In about 50\% cases, density gradient can be excluded, and none of the remaining objects could be confirmed. None of the observed planetary nebulae show the spectral index of 0.6 in the optically thick part of the spectrum, which is a value predicted for a shell containing strong radial density gradient. Radio spectra can be fitted with a model of prolate ellipsoidal shell, but also by a shell containing temperature variations in planetary nebulae. At least eight planetary nebulae show two component spectra, with one compact component showing much higher optical thickness than the other one. Unexpectedly, a group of planetary nebulae with lowest surface brightness show non-negligible optical thickness. Their emission comes from compact and dense structures, comprising only small part of total nebular mass.
\end{abstract}


\section{Introduction}

Low and intermediate mass stars ($ 1-8\, \textrm{M}_{\odot} $) play a vital role in
the chemical evolution of the Galaxy \citep{Marigo}. Up to 90\% of their initial 
mass returns to the interstellar medium, most of it during the asymptotic giant 
branch (AGB) phase of their evolution. 

After a long phase of steady hydrogen burning on the main sequence, hydrogen is
exhausted in the center of the star and nuclear burning proceeds in a shell
around a helium core. The star ascends the red giant branch in the
Hertzsprung-Russell diagram. Subsequently, helium ignites in the core. After
helium is exhausted in the core, helium burning proceeds in a shell on the top
of carbon-oxygen core. The star ascends the asymptotic giant branch. 

The main site of nucleosynthesis at the end of the AGB phase is the helium burning
shell, separated from the hydrogen burning shell by the intershell region.
Helium burning activates during the thermal pulse (helium flash), when enough
helium is accumulated in quiescent hydrogen burning. The material synthesized
during a thermal pulse is subsequently dredged up to the surface by the
convective envelope.

Thermal pulses trigger heavy mass loss. When the envelope mass
is reduced to ${\sim}10^{-3} \, M_{\odot}$, the star begins to evolve very
quickly to higher effective temperatures. Heavy mass loss terminates, and the slow and
dense AGB wind is being compressed by a fast wind from the central star, which
creates a central cavity. The density profile of the shell is further modified
by the passage of the ionization front. A planetary nebula (PN, plural PNe) becomes 
visible as a result of ionization and wind interaction.

\section{Radio emission of planetary nebulae}

Planetary nebulae are detectable in a broad range of the electromagnetic spectrum, from
UV or even X rays to radio frequencies. Continuum radio emission originates
from thermal free-free emission of electrons \citep{Pazderska,Chhetri}. Radio observations trace
all the ionized ejecta in the PN and give information on the nebular structure and 
physical parameters unaffected by extinction. This in turn can help to reveal the mass loss history of the star.

Spherically symmetric models cannot fit the observations. PNe show an excess of the 5\,GHz/1.4\,GHz 
flux ratio with respect to a model of a homogeneous spherical shell. \citet{Siodmiak} 
introduced a model consisting of two components to fit the observations.

On the other hand, \citet{Phillips} claimed that at least $10-20\%$ of PNe are
associated with strong density gradients, which can explain the observed 
5\,GHz/1.4\,GHz flux indices. For a shell with the density varying with 
radius $n_e(r) \propto r^{-2}$ the spectrum has a slope of ${\sim}0.6$ in the 
range of frequencies.

\citet{Gruenwald} were able to reproduce a range of spectral indices without 
introducing the initial density gradient, however, they analyzed the spectral 
indices limited to the frequency interval of 0.6 to 0.4 GHz.

The analysis using 5\,GHz/1.4\,GHz flux ratio by \citet{Siodmiak} and \citet{Phillips} did not lead to conclusive results. A prolate ellipsoidal shell was used by \citet{Masson} and \citet{Aaquist2} to fit radio images of PNe. The prolate ellipsoidal shell, viewed at different angles, can reproduce variety of observed nebular morphologies.

With the advent of many radio surveys, more data is available for different frequencies in addition to 5\,GHz and 1.4\,GHz fluxes, widely used in previous studies. We take advantage of recently published catalogs of radio fluxes covering low and high frequency region to further constrain the proposed models. New data helped us to constrain the slope of radio spectra at optically thick part, which can verify the existence of strong density gradients. We also attempted to fit prolate ellipsoidal shell model and models with electron temperature varying across the nebula to the observed spectral indices.

\section{Data}

\begin{table*}
\caption{Radio continuum observations used for modeling of nebular spectra.}
\label{table:0}
\centering
\begin{tabular}{c c c c c l}
\hline\hline
frequency [GHz] & sky coverage & sensitivity [mJy/beam] & beam size &  reference \\
\hline
0.15 & $\delta > -56^{\circ}$ & 25 & $20'' \textrm{or} \, 25'' \times 25''$ & \citet{Intema} \\
0.2 & $\delta < +30^{\circ}$ \&  $|b| > 10^{\circ}$ & 6.7& see reference &  \citet{Hurley-Walker}\\
0.325 & $\delta > +28^{\circ}$ & 30 & $54'' \times 54'' \mathrm{cosec} \, \delta$ & \citet{Rengelink} \\
0.330 &  $+42 < l < +92$, $|b| < 1.6^{\circ}$ & 10 & $1' \times 1' \mathrm{cosec} \, \delta$ & \citet{Taylor} \\
0.352  & $-26^{\circ} < \delta < -9^{\circ}$ & 18 & $54'' \times 54'' \mathrm{cosec} \, \delta$ & \citet{DeBreuck} \\
0.365 & $71.5^{\circ} > \delta > -35^{\circ}$ & 400 & see reference & \citet{Douglas} \\
0.843 & $\delta < -30^{\circ}, |b| >10^{\circ}$ & 18 (8  for $\delta < -50$) & $45'' \times 45'' \mathrm{cosec} \, \delta$ & \citet{Mauch} \\
0.843 & $|b| < -10^{\circ}, 245^{\circ} < l < 365^{\circ}$ & 10 (6 for $\delta < -50$) & $45'' \times 45'' \mathrm{cosec} \, \delta$ & \citet{Murphy} \\
1.4 & $\delta > -40^{\circ}$ & 2.5 & $45''$ & \citet{Condon} \\
1.4 & dedicated survey & see reference & $3.8'' \times 5.5''$ & \citet{Isaacman} \\
1.5 & dedicated survey & see reference & see reference & \citet{Zijlstra} \\
4.85 & $0^{\circ} < \delta < 75^{\circ}$ & 25 & $3.6' \times 3.4'$ & \citet{Gregory} \\
4.85 & $-87.5^{\circ} < \delta < 10^{\circ}$ & 35 & $4.2'$ & \citet{Griffith} \\
4.85 & dedicated survey & see reference & $2.8'' \times 1.5''$ & \citet{McConnell} \\
5 & dedicated survey & see reference & $4.9'' \times 6.5''$ & \citet{Isaacman} \\
5 & dedicated survey & see reference & see reference & \citet{Zijlstra} \\
5 & dedicated survey & see reference & $0.5''$ & \citet{Aaquist} \\
5 & $\delta < 27^{\circ}$ & see reference & $4.5''$ & \citet{Milne75} \\
5 & $\delta < 0^{\circ}$ & 40 & see reference & \citet{Murphy2} \\
8 & $\delta < 0^{\circ}$ & 40 & see reference & \citet{Murphy2} \\
8.6 & dedicated survey & see reference & $1.3'' \times 0.9''$ & \citet{McConnell} \\
14.7 & $\delta < 27^{\circ}$ & see reference & $2.1''$ & \citet{Milne82} \\
15 & dedicated survey & see reference & see reference & \citet{Zijlstra} \\
15 & dedicated survey & see reference & $0.1''$ & \citet{Aaquist2} \\
16 & $|b| \leq 5, 170^{\circ} > l > 76^{\circ}$ & 16 & $3'$ & \citet{Perrott} \\
20 & $\delta < 0^{\circ}$ & 40 & see reference & \citet{Murphy2} \\
30 & $\delta > -15^{\circ}$ & 20 & $1.2'$ & \citet{Pazderska} \\
30 & all sky & 427 & $32'$ & \citet{Planck} \\
31 & dedicated survey & 20 & $45.2''$ & \citet{Casassus} \\
43 & dedicated survey & 50 & $54''$ & \citet{Umana} \\
44 & all sky  & 692 & $27'$ & \citet{Planck} \\
70 & all sky & 501 & $13.2'$ & \citet{Planck} \\
100 & all sky & 269 & $9.7'$ & \citet{Planck} \\
143 & all sky & 177 & $7.2'$ & \citet{Planck} \\
217 & all sky & 152 & $4.9'$ & \citet{Planck} \\
353 & all sky & 304 & $4.9'$ & \citet{Planck} \\
353 & dedicated survey & 71 & $14''$ & \citet{Scuba} \\
\hline
\end{tabular}
\end{table*}

We searched for archival radio observations of PNe having at least three observations
at different frequencies in total, including at least one observation in low or
high frequency surveys to ensure wide coverage in frequencies.

Low frequency surveys between 150 and 352\,MHz were carried on the Westerbork Synthesis Radio Telescope \citep{Rengelink,Taylor,DeBreuck}, Giant Metrewave Radio Telescope \citep[TGSS survey,][]{Intema}, Murchison Widefield Array \citep[GLEAM survey,][]{Hurley-Walker}, and the Texas interferometer \citep{Douglas}. High frequency surveys were carried by \citet{Pazderska}, \citet{Casassus}, \citet{Umana}, and the \citet{Planck}. For intermediate frequencies we used data from
surveys which derived fluxes with accuracy ${\leq}10\%$. The data are summarized in Table \ref{table:0}. We used SIMBAD database \citep{Wenger} to identify spurious background sources.

We used nebular diameters from \citet{Frew}. Radio diameters available in the literature become unreliable for small nebulae are less certain than optical diameters \citep{Siodmiak}. \citet{Frew} measured diameters for the emission region at the 10\% level of the surface brightness, which should represent the shell contributing most of the flux in radio. In the case of non-spherical PNe, we adopted the geometric mean of the two axes.
 
\section{Modeling of radio spectra} \label{Sec4}

For an isothermal nebula the radio flux $S_{\nu}$ is given by 

\begin{equation} \label{flux} S_{\nu}=\frac{2\nu{^2} kT_e}{c^2}(1-e^{-\tau_{\nu}})\Omega \end{equation}
and optical thickness of the nebula $\tau_{\nu}$

\begin{equation} \label{tau} \tau_{\nu} = 5.44 \times 10^{-2} T_e^{-1.5} \nu^{-2} g_{ff}(\nu , T_e) \int n(H^+) n_e dS \end{equation}
where $T_e$ is electron temperature, $\Omega$ is the solid angle, $n(H^+)$ and $n_e$ are proton and electron density, respectively. The Gaunt factor is given by $g_{ff}$. Emission measure $EM = \int n_e \, n(H^+) \, dS$ is integrated along the line of sight. \citet{vanHoof} provided accurate values of the Gaunt factor $g_{ff}$.

We used five different models to fit observations: a homogeneous sphere, a cylindrical shell seen along its axis, a spherical shell with uniform density distribution, a spherical shell containing radial density gradient, and a prolate ellipsoidal shell. We also checked if $T_e$ variations in a PN could fit the observations.

\subsection{Model A: homogeneous sphere}

As a first approximation, we fitted a model of a homogeneous sphere. There are three free parameters in equations \ref{flux} and \ref{tau}: $\Omega$, $T_e$, and $EM$. $EM$ primarily determines the turnover frequency $\nu$ of the PN spectrum for which $\tau_{\nu}{\sim}1$. The optically thin part of the spectrum has a spectral index of about $-0.1$. The absolute flux is proportional to $EM$. The optically thick part of the spectrum has a spectral index of 2. The absolute flux is proportional to $T_e$. Additionally, $\Omega$ scales the absolute flux in the whole range of the spectrum. $EM$ and $T_e$, can be measured unambiguously when the radio spectrum covers both optically thick and thin emission. This is the case for most of PNe studied by us, which show decline of the flux due to increasing optical thickness at lowest frequencies.

Model A in most cases gave electron temperatures significantly lower than ${\sim}10^4 \, K$, expected for photo ionized nebula in equilibrium (Fig. \ref{fit}). Such low temperatures contradict optical observations and physics of PNe. Higher $T_e$, however, would overestimate the flux in optically thick region in model A. Thus, model A is not relevant for most of PNe. 

\newpage
\subsection{Model B: homogeneous shell}

In order to derive more physically plausible models, we fixed $T_e$ to the value derived from optical spectra $T_{opt}$ \citep{Cahn}. We used a model of uniform sphere, but with a central cavity defined by the ratio of the inner to outer radius $\eta_B = R_{in} / R$. The fitted parameters in model B were $\eta_B$ and $EM$. 

The shell model with ${\eta}_B{\approx}1$ shows strong limb brightening. The $EM$ reaches the highest value at the projected distance of $R_{in}$ from the center of the PN and is low elsewhere, which may reproduce the observations. Model B with ${\eta}_B{\approx}1$ would require a very high density and would produce ring shaped PNe. On the other side, model B with ${\eta}_B{\approx}0$ is similar to model A.

\begin{figure}
\centering
\includegraphics[width=\hsize]{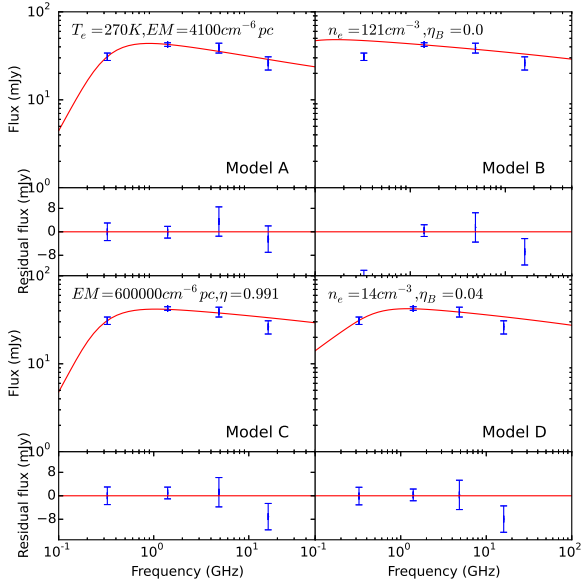}
\includegraphics[width=\hsize]{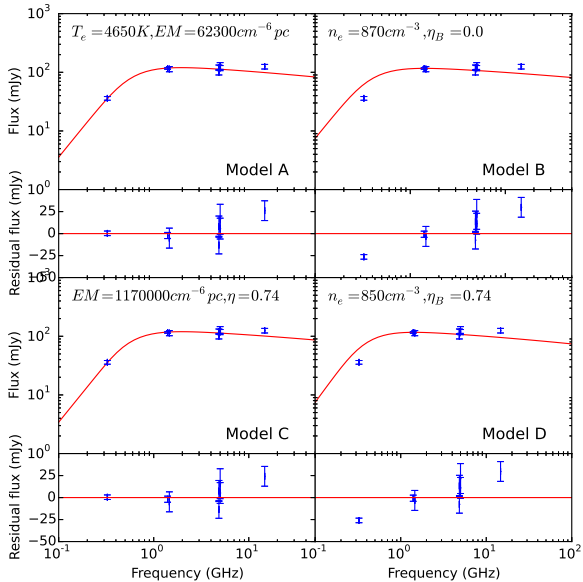}
\caption{Four fits to the radio spectra of PN NGC\,6842 (top) and IC\,4634 (bottom) using models A, B, C, and D. The derived parameters for each model are shown. Lower panels show the residuals of the fits.}
\label{fit}
\end{figure}

The fitted $EM$ reproduced well the flux density but was too low to reproduce the turnover frequency in many cases (Fig. \ref{fit}). Higher $EM$ would fit the turnover frequency better, but also would overestimate the absolute flux in optically thin part of the spectrum. Thus, model B is not relevant for most of PNe.

\subsection{Model C: homogeneous cylindrical shell}

We introduced the $(1-\eta^2)$ factor in equation \ref{flux} so that we could increase $EM$ to achieve a good fit of the turnover region and simultaneously scale down the absolute flux:

\begin{equation} \label{cyl} S_{\nu}=\frac{2\nu{^2} kT_e}{c^2}(1-e^{-\tau_{\nu}})\Omega (1 - \eta^2) \end{equation}
where fitted parameters are $\eta$ and $EM$. The $\eta$ parameter is defined as the ratio of the {\it projected} inner to outer radius $R_{in} / R$.

Model C reproduced most of the spectra very well. It can be interpreted as a cylindrical shell seen along its axis, with the ratio of the internal to external radii defined by $\eta$. Model C can be relevant as a first approximation for some PNe, including the well known NGC\,6720, which shows a disk seen pole on \citep{Odell}. Assuming random orientation of PNe in space, this model cannot account for most of PNe.

\begin{figure}
   \centering
   \includegraphics[width=\hsize]{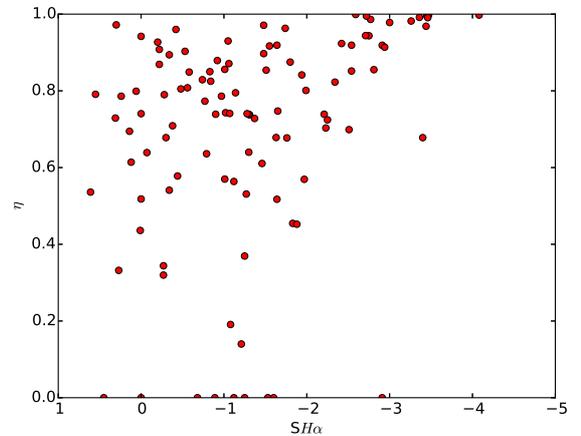}
   \caption{Diagram showing $\eta$ vs $\textrm{H}\alpha$ surface brightness.}
              \label{shvseta}%
    \end{figure}

However, we find this model useful, as it is sensitive to the most optically thick part of the PN. The $\eta$ parameter gives information on surface brightness distribution of a PN. PNe with $\eta{\approx}0$ can be fit equally well using model A. Almost all PNe require $\eta > 0$ (Fig. \ref{shvseta}). Low surface brightness PNe need $\eta$ very close to 1. High value of $\eta$ indicates, that a small region with high $EM$ dominates in radio.

   \begin{figure}
   \centering
   \includegraphics[width=\hsize]{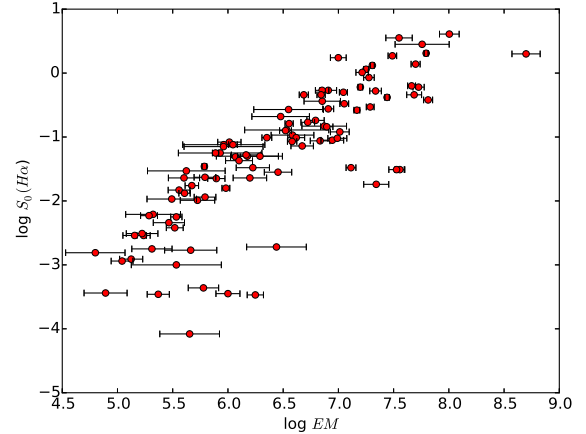}
   \caption{Emission measure vs $\textrm{H}{\alpha}$ surface brightness in observed PNe.}
              \label{emt}%
    \end{figure}

   \begin{figure*}
   \centering
   \includegraphics[width=\hsize]{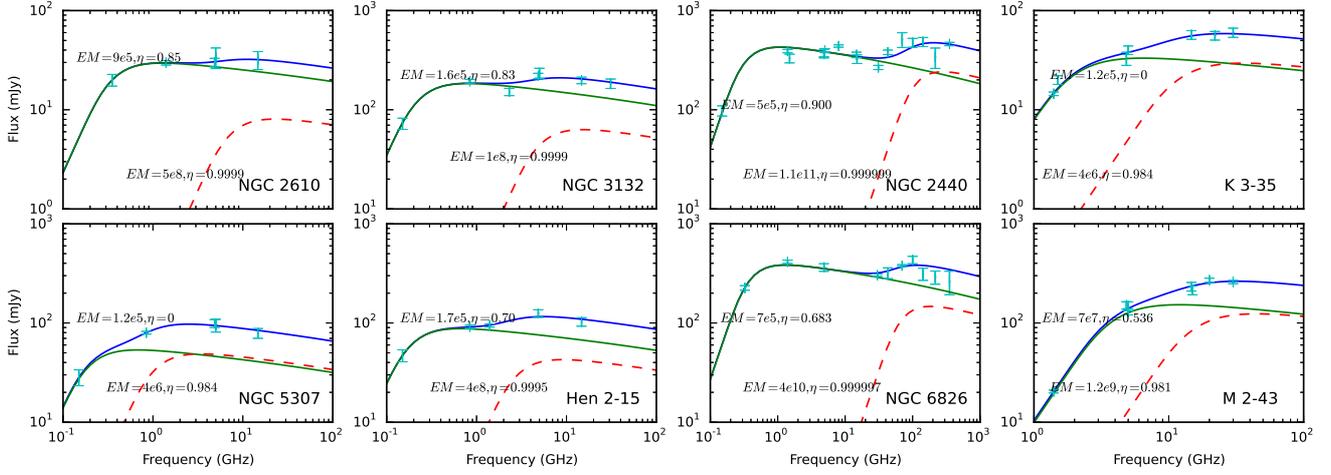}
   \caption{PNe showing two components in their spectra. Dashed red and green lines show separate components, blue line shows the sum of both components.}
               \label{twocomp}%
    \end{figure*}

   \begin{figure}
   \centering
   \includegraphics[width=\hsize]{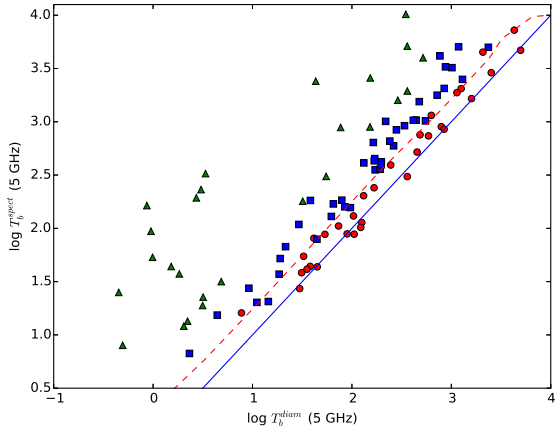}
   \caption{Brightness temperatures determined from measured diameter and flux and from radio spectra. PNe are split into three groups: $\eta < 0.5$ (red circles), $0.5 < \eta < 0.9$ (blue squares), and $\eta > 0.9$ (green triangles). Dashed red curve shows Model E for $a/b=3$, $t/b=0.1$, and $i=0^{\circ}$, and blue solid curve shows $T_b^{diam} = T_b^{spec}$ relation.}
              \label{tb}%
    \end{figure}

   \begin{figure}
   \centering
   \includegraphics[width=\hsize]{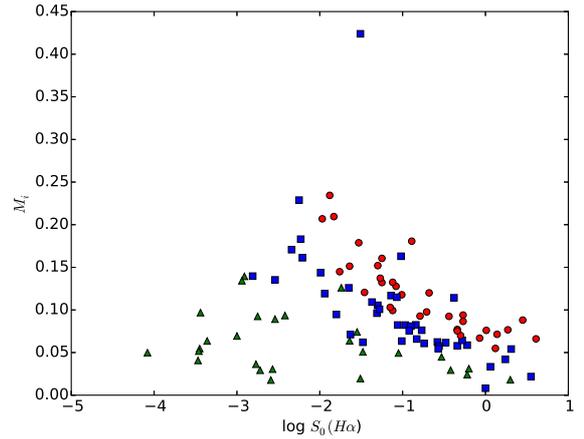}
   \caption{Ionized mass vs dereddened $H \alpha$ surface brightness.}
              \label{mion}%
    \end{figure}

Fig. \ref{emt} presents the relation between emission measure derived in model C and dereddened $\mathrm{H}{\alpha}$ surface brightness \citep{Frew}. As expected, $EM$ decreases with 
$S_0(\mathrm{H}\alpha)$, but unexpectedly this tendency is not held for PNe with the lowest surface brightnesses.

We also used a model consisting of two components, fitting two different values of $EM$ and $\eta$. In eight cases two components improves the fit significantly (Fig. \ref{twocomp}). One of the components has much higher turnover frequency and a flux comparable to the second component. This indicates much higher $EM$ of the order of $10^8 \, \mathrm cm^{-6} \, pc$, which could indicate a compact and dense (possibly cool) component in a PN. 

In the appendix we list $EM$ and $\eta$ fitted to equation \ref{cyl}. In addition, we present optical depth at 5\,GHz, brightness temperature at 5\,GHz derived from fitting the spectrum $T_b^{spect} = T_e \times (1 - e^{-{\tau}_{5GHz}})$, and from $T_b^{diam} = 73.87 \, F_{5GHz} / {\Theta}^2$, where $\Theta$ is the diameter of the PN measured in arcsec, and $F_{5GHz}$ is flux at 5\,GHz measured in mJy.
 
$T_b^{spect}$ is systemically higher than $T_b^{diam}$ (Fig. \ref{tb}). For uniform brightness distribution, corresponding to $\eta = 0$, $T_b^{spect} = T_B^{diam}$. For non uniform brightness distribution, $T_b^{spect} > T_B^{diam}$. The ratio of $T_b^{spect} / T_B^{diam}$ in Model C simply depends on $\eta$: $T_b^{spect} / T_B^{diam} = 1 - \eta^2$.

We derived ionized masses using distances derived by \citet{Frew} and emission measures from radio spectra using model C. Ionized masses increase with decreasing surface brightness down to $\log(S_o(\mathrm{H} {\alpha} )) = -2$ (Fig. \ref{mion}) which reflects the progressing ionization in the nebulae \citep{Frew}. For lower surface brightness most of the observed PNe have $\eta > 0.9$. Fig. \ref{mion} includes only PNe with reliable determination of $EM$, so low surface brightness PNe which are optically thin in radio are not included.

\subsection{Model D: a shell with a density gradient $\beta = 2$}

We also used the modified model of power-law distribution by \citet{Olnon} for $n_e = n_o (r/(\eta_B \times R))^{-\beta}$ for $\eta_B \times R > r > R$, $n = 0$ for $r \leq \eta_B \times R$, and $\beta = 2$. This density distribution can result from a steady-state wind lasting for a finite time interval. In this model we fitted $\eta$ and $n_o$.

Model D results in the flat spectrum in an optically thin part and in a spectral index of 0.6 in partially optically thick part, and an index of 2 in the optically thick part. The range of spectrum where spectral index is 0.6 depends on the value of $\eta_B$: the lower $\eta_B$, the larger range of frequencies where the spectral index of 0.6 is observed.

\subsection{Model E: prolate ellipsoidal shell}

Surface radio brightness distribution suggests that a model of a prolate ellipsoidal shell can be representative for most of PNe \citep{Masson,Aaquist2}. The inner and outer shell are defined as ellipsoids with minor and major axes of $(a,b)$ and $(a+t,b+t)$. Electron density depends on the incident flux, so that $n_e^2 \times t$ is inversely proportional to the inner radius.

Due to large parameter space (inclinaton angle $i$, $a/b$ ratio, $t$, $n_e$), we did not use Model E to fit individual PNe, but to fit spectral indices and brightness temperatures.

\subsection{Model F: temperature stratification}

We also explored the influence of temperature stratification on the radio spectra of PNe. We assumed a spherical model with $T_e$ of 2500\,K in the central region and 10000\,K elsewhere, and a model with $T_e$ of 1000\,K in the outer 20\% radius and 10000\,K inside. The temperature stratification is set arbitrarily. Again, we did not use this model for individual PNe, but rather to fit all PNe in the brightness temperature -- spectral index diagrams.

\section{Discussion}

\subsection{Spectral indices}

\begin{figure}
   \centering
   \includegraphics[width=\hsize]{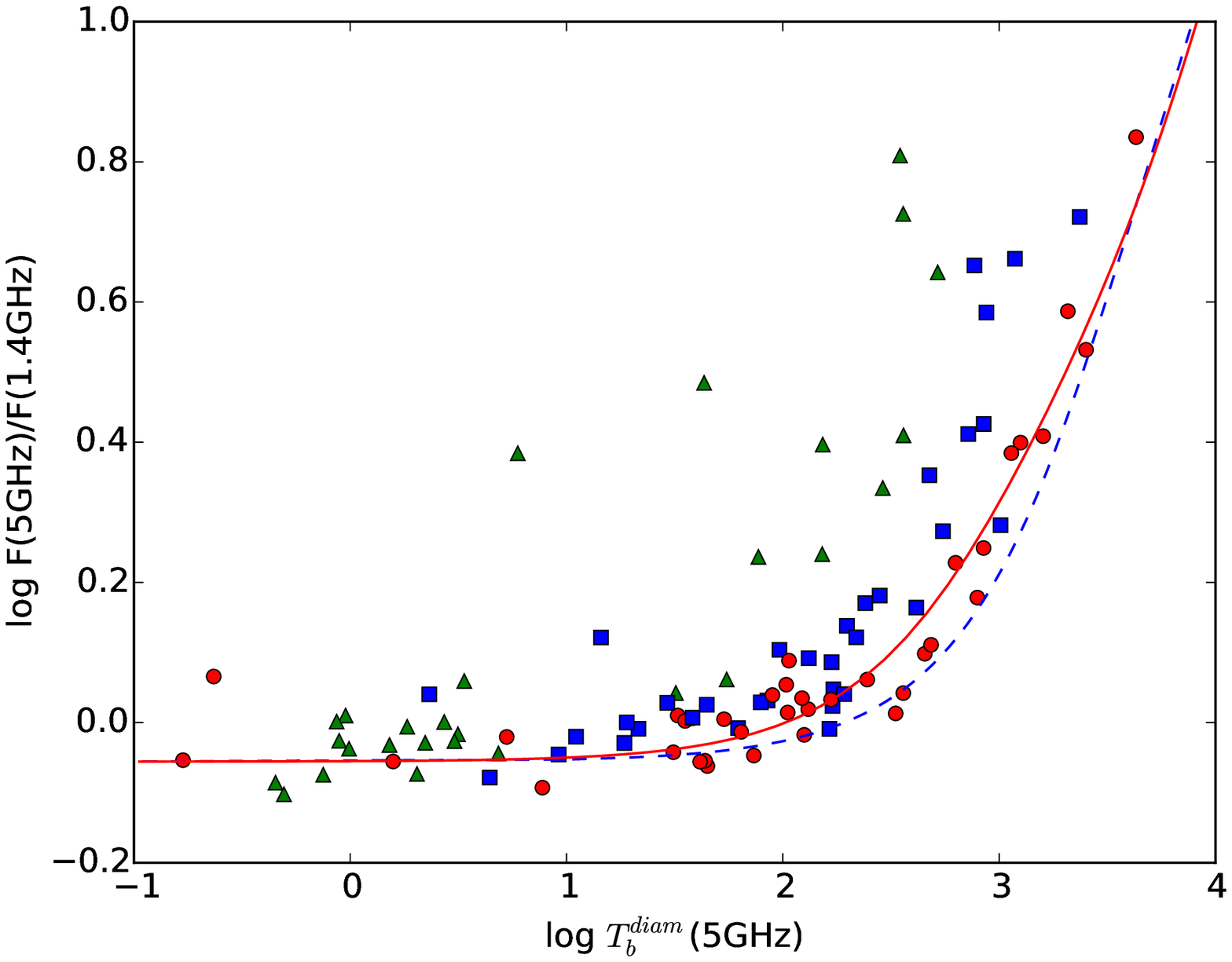}
   \includegraphics[width=\hsize]{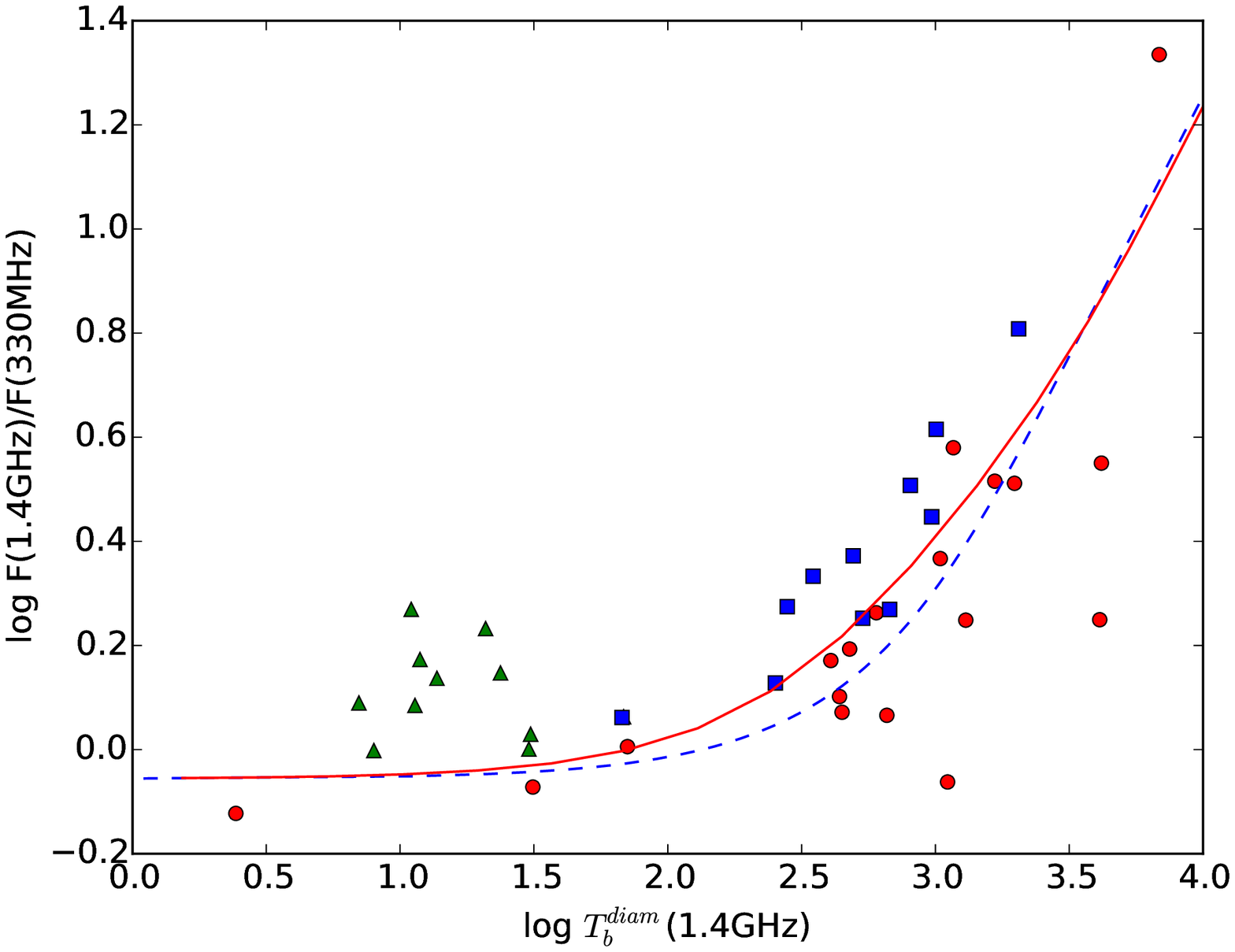}
   \caption{The observed spectral indices vs brightness temperature for PNe and model A for $T_e = 10000\,K$ (blue line). Red curves show Model E for $a/b=3$, ${t/b}=0.1$, and $i=0^{\circ}$. PNe are marked with the symbols as in Fig. \ref{tb}.}
               \label{indices}%
    \end{figure}

\begin{figure}
   \centering
   \includegraphics[width=\hsize]{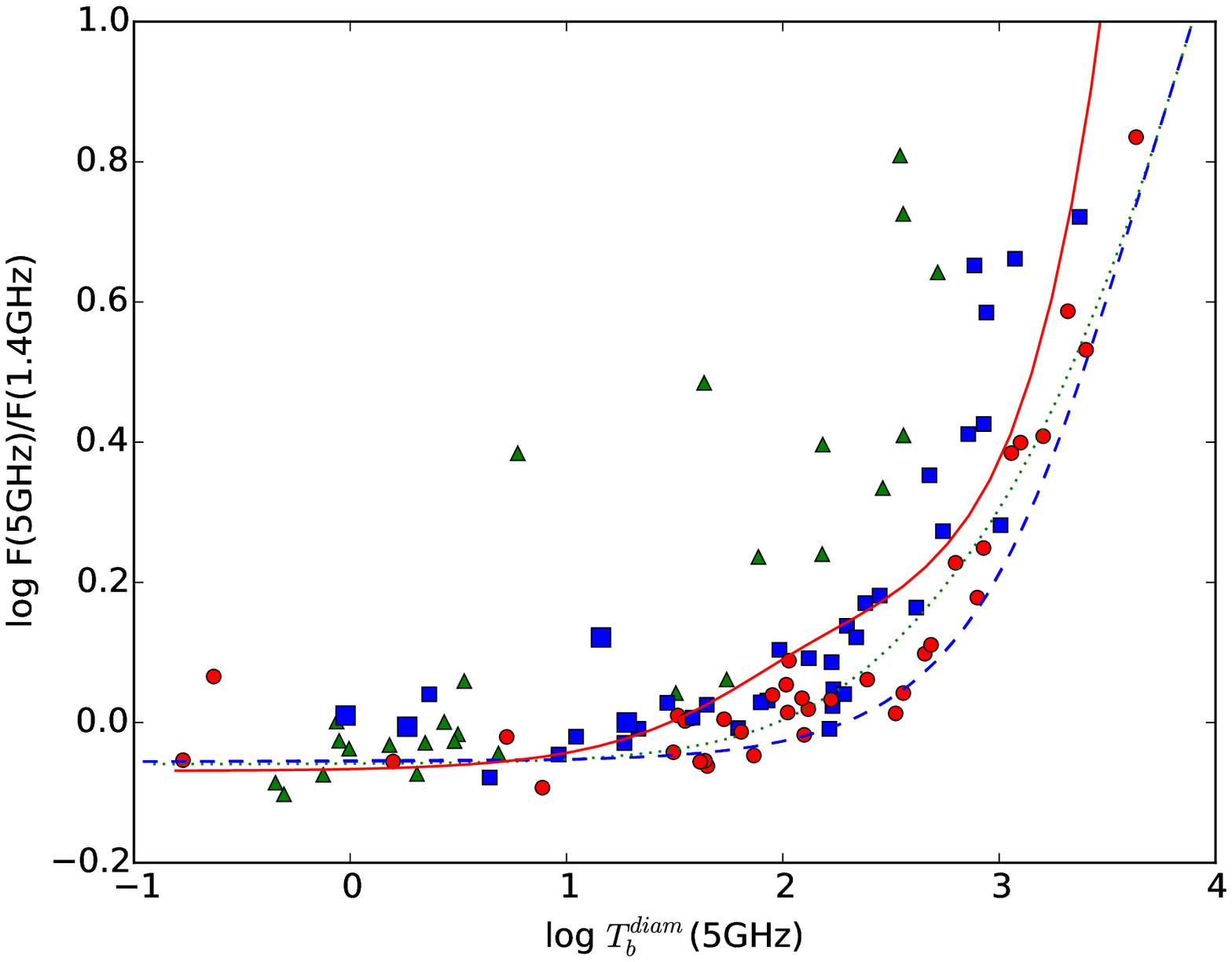}
   \includegraphics[width=\hsize]{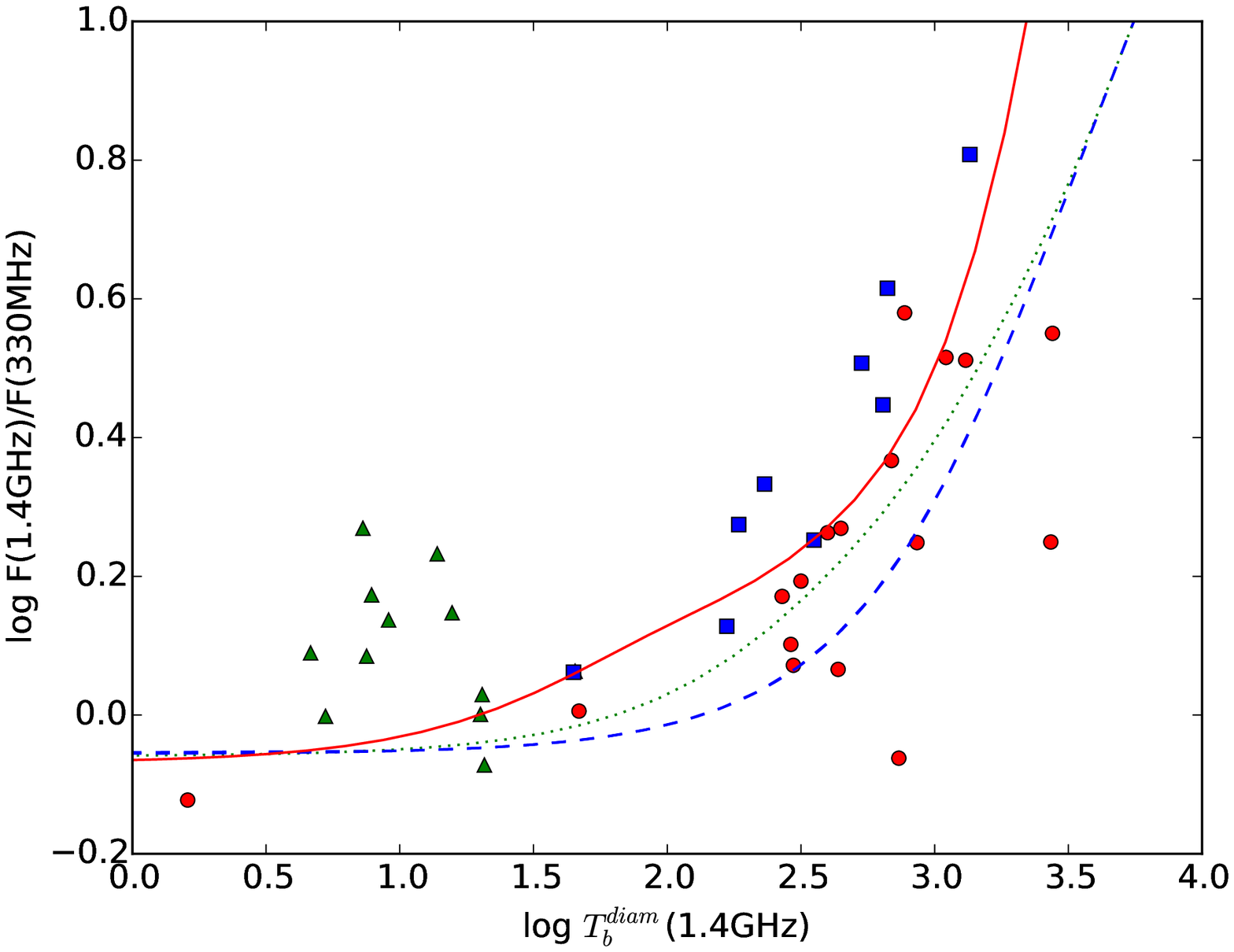}
   \caption{The observed spectral indices vs brightness temperature for PNe and model A for $T_e = 10000\,\mathrm{K}$ (blue dashed line). Dotted curve shows model with inner region of lower temperature, while red solid curve shows model with outer region of lower temperature. PNe are marked with the symbols as in Fig. \ref{tb}.}
               \label{indicest}%
    \end{figure}

\begin{figure*}
   \centering
   \includegraphics[width=\hsize]{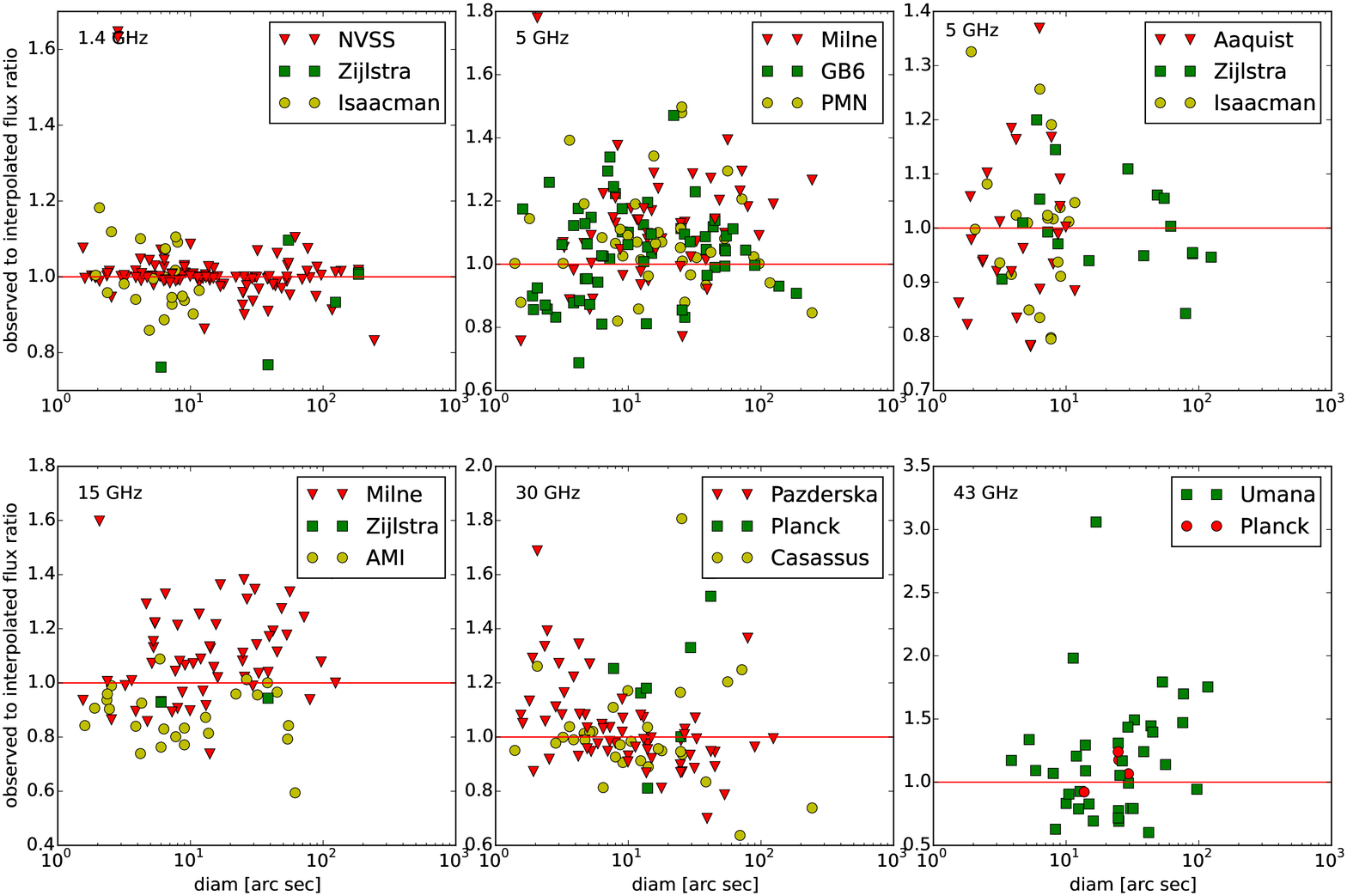}
   \caption{The ratio of the observed flux to the flux derived for model C vs PN diameter for different frequencies. For 5\,GHz, single dish and interferometric surveys are shown separately.}
              \label{excess}%
    \end{figure*}

A diagram of spectral indices vs $T_b$ reveal an excess of the {observed} flux ratios
with respect to model A and B (Fig. \ref{indices}, \ref{indicest}).
The excess does not depend on the nebular diameter nor on the instrument used (Fig.
\ref{excess}), except for \citet{Milne75}, where the fluxes
appear to be correlated with nebular size. Data by \citet{Milne82} at higher 
frequencies show a similar problem. The data published by \citet{Perrott} appear 
to be underestimated. Apart from that the different instruments used does not seem to affect
the observed indices.

In some cases, the excess is visible only at 5\,GHz, e.g. in IC\,3568, IC\,5217, IC\,289, or NGC\,6072. Out of them, only IC\,5217 shows a weak, confusing source within the primary beam of the survey of 3.5 arcmin \citep{Gregory}, which might be responsible for the excess of the 5\,GHz/1.4\,GHz ratio. However, the excess of the 5\,GHz/1.4\,GHz ratio does not depend on the PNe diameters, and thus cannot be affected by the presence of confusing sources in the single dish surveys. Also, 5\,GHz excess was reported by \citet{Siodmiak}, who used only interferometric fluxes.

\subsection{Flux evolution}

One possible explanation for the 5\,GHz excess could be the evolution of the radio fluxes due to nebular expansion and change of the ionizing flux from the central star. 5\,GHz fluxes used in our paper are predominantly from the single dish surveys \citep{Gregory,Griffith} carried out in the period $1986-1987$ and 1990. 1.4\,GHz observations were taken predominately from the NRAO VLA Sky Survey (NVSS) between 1993 and 1996. Since 1.4\,GHz flux should increase with respect to the 5\,GHz thorough the PN evolution, the time gap between the 5\,GHz surveys and NVSS should rather result in lowering the 5\,GHz/1.4\,GHz ratio.

\citet{Zijlstra2} reported variability of radio flux of NGC\,7027 due to expansion of the nebula (increase of $0.25\% \, yr^{-1}$) and decreasing number of ionizing photons (change of $-0.14\% \, yr^{-1}$). Such small changes would be difficult to detect with the precision of about 10\% and relatively short time span between observations. However, flux evolution depends on the age of the nebula and pace of evolution of central star.

We checked the possible flux evolution using observations at the same frequency made for the same PN in different epochs, if available. We selected sources which showed more than $3 \sigma$ deviation in measurements at the same frequency. In most cases, the flux measurements did not show significant changes. We rejected sources in which the possible reason for discrepancy was systemic flux error at one epoch, which did not show the expected flux changes at other frequencies and the overall flux spectrum did not show any signatures of variability, e.g. in Hu\,1-2.

The variability of radio flux in IC\,4997 has already been studied by \citet{Miranda}. They showed year-to-year changes of the morphology of PN IC\,4997 in radio due to interaction of the collimated stellar wind with the outer shell. In the archival data, the flux at optically thin part dropped from 100\,mJy around 1980 down to 50\,mJy about 2000. More recent data at higher frequencies show possible increase from 80 to 110\,mJy between $2001-2002$ \citep{Casassus} and $2005-2007$ \citep{Pazderska}, respectively. However, the data are too sparse to fit the spectrum for each epoch separately.

Another case is M\,2-2. This is a relatively compact PN with the diameter of only 6 arcsec. \citet{Zijlstra} observed higher values at 5\,GHz and 15\,GHz, and lower flux at 1.5\,GHz than more recent observations. This would imply the lowering flux in continuum and increasing flux in optically thick part of the spectrum, in qualitative agreement with NGC\,7027 \citep{Zijlstra2}. Another possible variable is M\,1-40, showing a decreasing trend over time. We would need more data to analyze the flux variability in more details. Most of PNe did not show any variability.

\subsection{Radial density gradient}

\begin{figure*}
   \centering
   \includegraphics[width=\hsize]{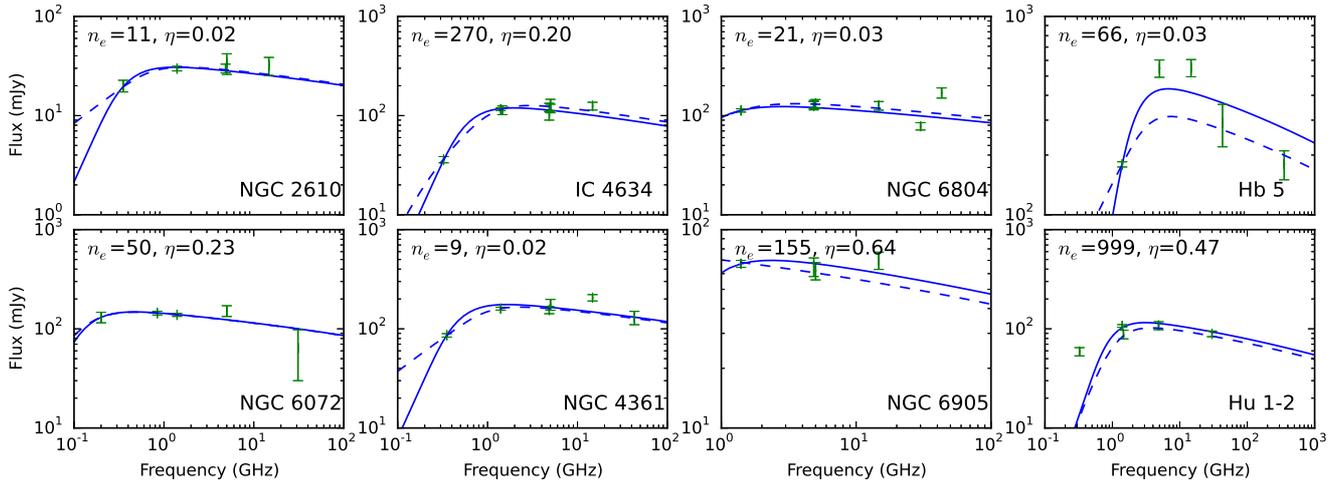}
   \caption{Spectra of PNe claimed to have strong density gradient. Solid line shows fit of model C with the parameters given in the appendix, and dashed line shows fit of model D with the parameters indicated.} 
               \label{dgcandidates}%
    \end{figure*}

A spectral index of 0.6 is an indication of a strong radial density gradient ($\beta = 2$). The spectral index of 0.6 should be relatively easy to confirm due to a relatively high flux at low frequencies. \citet{Phillips} and \citet{Phillips2} list candidates for strong density gradient. Eight of them are in common with our sample (Fig. \ref{dgcandidates}). The objects were well fitted using model C except for Hu\,1-2. However, the diameter of Hu\,1-2 reported by \citet{Frew} apparently refers to the brightest, central region of the PN, while apparently much larger region of the nebula contributes to the observed radio flux \citep{Fang}. With larger diameter we received much better fit.

Most of PNe could still be fitted with model D, but it yields worse or unrealistic fit compared to model C in about 50\% of PNe. For a small value of $\eta_B$ model B gives artificially small densities $n_o$ or cannot fit the slope steeper than 0.6 in the optically thick part of the spectrum (e.g. Fig. \ref{fit}, \ref{dgcandidates}). Small ${\eta}_B$ would also result in the surface brightness quickly fading with the distance from the center of PN, and this is not confirmed in radio images \citep{Aaquist2}. 

Alternatively, the observed PNe may only have thin shells. Then, the spectral index of 0.6 is observed only in a very limited range of frequencies and is difficult to confirm. However, Model D is almost equivalent to model B for large $\eta_B$, and cannot fit the observations of most of PNe for the same reason, described in Section 4.2.

\subsection{Prolate ellipsoidal shell and temperature variations}

Model E explains well the observed spectral indices of many PNe with a typical value of a/b of about 3. Alternatively, observed spectral indices could result from temperature variations. Lower temperature in the outer layer results also in lower spectral index with respect to a homogeneous model in the optically thick part of the spectrum. It would eventually produce a spectral index steeper than 2.

An upper limit from the TGSS survey indicates that some PNe show steeper spectral index than 2 in the optically thick part: NGC\,6826, NGC\,6543, and NGC\,6818. Optically thick radio emission represents the emitting region at $\tau{\sim}2/3$, whereas optically thin emission depends on volume averaged electron temperature. Thus, at 5\,GHz the excess can be interpreted as the temperature drop in the outermost regions of PN. The drop of temperature in the outer regions is not confirmed by spatially resolved spectroscopy of PNe \citep{Sandin}.

Model F with the temperature of 1000\,K in the central region of the shell gives more flux in the optically thin region, as the emissivity of lower temperature region is higher. The optically thick spectrum is the same for the two models as radio emission originates from the outermost region of the shell. The region with lower temperature was indicated by \citet{Tsamis} and is one of the alternative explanations for the abundance discrepancy problem in PNe. \citet{Garcia} showed that emission originating from optical recombination lines is more centrally concentrated than emission from collisionally excited lines.

\subsection{Low surface brightness PNe}

Unexpectedly, a group of PNe with low $T_b^{diam}$ shows non-negligible optical thickness at lowest frequencies (Figs. \ref{indices},\ref{indicest}). Model C gives $\eta > 0.9$, which indicates that a small part of the projected area of the PN with high $EM$ contribute most of the radio flux. However, NVSS and other catalogs give the radii comparable to that seen in optical. This suggests that they form a cylindrical shell observed along its axis, which causes strong limb brightening, or contain a compact component, which contribution in flux is comparable to the rest of PN. Relatively low $M_i$ indicate that these structures comprise only part of the nebular ionized mass.

Figure \ref{shvseta} shows, that for low surface brightness PNe $S_0(\mathrm{H}\alpha) < -2$ only those with large $\eta$ parameter are observed. According to \citet{Aaquist2}, morphology of PNe does not change between low and high surface brightness PNe. This indicates, that low surface brightness PNe are detected only when they contain a small region with high column density.

This group contains bipolar or quadrupolar PNe NGC\,650-51, NGC\,4361, NGC\,6302, or M\,1-75. Also, three PNe classified as round, NGC\,2610, Abell 53, and NGC\,6842, belong to this group, indicating, that their spatial structure is more complex than spherical.

\section{Summary}

In conclusion, we confirm that homogeneous models cannot fit the observed radio spectra and brightness temperatures of most of PNe. We do not find an evidence for a strong radial density gradient in PNe. We found a sample of PNe showing two component spectra, but they cannot account for the observed 5\,GHz to 1.4\,GHz indices. A model of prolate ellipsoidal shell or temperature variations explain very well the observed spectral indices with except of low surface brightness PNe. Part of low surface brightness PNe are only observed in radio due to their specific morphology. Further observations at low frequencies having better sensitivity and resolution, e.g. Low Frequency Array \citep{vanHaarlem} will better constrain the spectral index in optically thick part of the spectra and geometry of PNe.

\section*{Acknowledgements}

We thank to the Ministry of Science and Higher Education (MSHE) of Republic of Poland for granting funds for Polish contribution to the International LOFAR Telescope (MSHE decision no DIR/WK//2016/05) and for maintenance of the LOFAR PL- 612 Baldy (MSHE decision no 220815/E-383/SPUB/2016/2). We gratefully acknowledge financial support from National Science Centre, Poland, grant No. 2016/23/B/ST9/01653. This research has made use of the SIMBAD database, operated at CDS, Strasbourg, France.

\appendix

\section{Emission measures and $\eta$ values derived for Model C}

\onecolumn
\begin{longtable}{l c c c c c c c c c}        
\caption{Modeled temperatures and emission measures.} \\
\hline
\hline
Name & RA (J2000) & DEC (J2000) & $EM [cm^{-6} pc]$ & error EM & $\eta$ & error $\eta$ & $\tau_{5GHz}$ & $T_b^{spec}$ & $T_b^{diam}$  \\
\hline
NGC 40		&	00 13 01.010 	&	+72 31 19.09 	&	3.476E+5	&	6.943E+4	&	0.7283	&	0.0563	&	0.00338	&	  37.158	&	  18.260\\
Hu 1-1		&	00 28 15.435 	&	+55 57 54.48 	&	4.206E+5	&	3.375E+5	&	0.0000	&		&	0.00360	&	  43.508	&	  45.537\\
NGC 246		&	00 47 03.338 	&	-11 52 18.94 	&	2.174E+3	&	2.008E+5	&	0.0000	&		&	0.00001	&	   0.205	&	   0.214\\
NGC 650-51	&	01 42 19.69 	&	+51 34 31.7 	&	2.315E+5	&	5.031E+4	&	0.9907	&	0.0019	&	0.00234	&	  24.995	&	   0.485\\
IC 1747		&	01 57 35.734 	&	+63 19 18.33 	&	4.005E+5	&	1.284E+5	&	0.5174	&	0.2177	&	0.00437	&	  44.054	&	  33.766\\
IC 289		&	03 10 19.30 	&	+61 19 01.0 	&	8.199E+4	&	9.264E+4	&	0.6561	&	0.4832	&	0.00050	&	   7.792	&	   4.645\\
M 1-4		&	03 41 43.439 	&	+52 16 59.85 	&	3.046E+6	&	1.756E+6	&	0.0000	&		&	0.02419	&	 305.881	&	 320.152\\
IC 2003		&	03 56 22.027 	&	+33 52 29.27 	&	4.048E+5	&	1.118E+6	&	0.0000	&		&	0.00347	&	  41.869	&	  43.822\\
NGC 1501	&	04 06 59.39 	&	+60 55 14.4 	&	3.304E+5	&	5.757E+4	&	0.9234	&	0.0126	&	0.00210	&	  31.671	&	   4.885\\
NGC 1514	&	04 09 16.985 	&	+30 46 33.47 	&	7.418E+4	&	3.716E+4	&	0.9669	&	0.0163	&	0.00074	&	   7.989	&	   0.544\\
M 2-2		&	04 13 15.043 	&	+56 56 58.12 	&	8.573E+5	&	8.093E+5	&	0.0000	&		&	0.00726	&	  88.263	&	  92.380\\
NGC 1535	&	04 14 15.765 	&	-12 44 21.90 	&	1.927E+5	&	3.384E+4	&	0.7032	&	0.0604	&	0.00173	&	  20.182	&	  10.678\\
IC 418		&	05 27 28.203 	&	-12 41 50.26 	&	8.455E+6	&	1.807E+6	&	0.3525	&	0.2460	&	0.09743	&	 900.472	&	 825.399\\
NGC 2022	&	05 42 06.20 	&	+09 05 10.3 	&	1.671E+5	&	5.508E+4	&	0.6989	&	0.1180	&	0.00108	&	  16.099	&	   8.620\\
M 1-5		&	05 46 50.01 	&	+24 22 02.8 	&	7.838E+6	&	1.145E+6	&	0.3964	&	0.1215	&	0.09837	&	 852.525	&	 752.131\\
IC 2149		&	05 56 23.901 	&	+46 06 17.19 	&	1.041E+6	&	1.975E+5	&	0.1948	&	0.4433	&	0.01107	&	 113.344	&	 114.128\\
IC 2165		&	06 21 42.775 	&	-12 59 13.96 	&	4.652E+6	&	1.113E+6	&	0.7950	&	0.0501	&	0.03305	&	 451.917	&	 174.056\\
J 900	&	06 25 57.237 	&	+17 47 27.53 	&	1.973E+6	&	8.423E+5	&	0.6452	&	0.1800	&	0.01690	&	 202.745	&	 123.857\\
M 1-6		&	06 35 45.126 	&	-00 05 37.36 	&	1.108E+7	&	9.299E+5	&	0.6784	&	0.0240	&	0.11939	&	1147.872	&	 648.450\\
NGC 2371	&	07 25 34.68 	&	+29 29 26.4 	&	1.374E+5	&	4.887E+4	&	0.9216	&	0.0285	&	0.00096	&	  13.496	&	   2.127\\
NGC 2392	&	07 29 10.765 	&	+20 54 42.48 	&	2.913E+5	&	1.030E+5	&	0.8226	&	0.0678	&	0.00174	&	  27.486	&	   9.301\\
NGC 2438	&	07 41 50.51 	&	-14 44 07.7 	&	1.667E+6	&	2.008E+6	&	0.9975	&	0.0028	&	0.01643	&	 177.596	&	   0.926\\
NGC 2440	&	07 41 54.91 	&	-18 12 29.7 	&	5.303E+5	&	2.003E+5	&	0.8018	&	0.0829	&	0.00373	&	  52.150	&	  19.491\\
Hen 2-9		&	08 28 27.924 	&	-39 23 39.92 	&	4.864E+6	&	4.600E+5	&	0.4135	&	0.0560	&	0.05240	&	 520.713	&	 451.809\\
NGC 2610	&	08 33 23.40 	&	-16 08 57.5 	&	1.042E+6	&	2.562E+5	&	0.9955	&	0.0010	&	0.00530	&	  94.097	&	   0.885\\
Hen 2-11	&	08 37 08.10 	&	-39 25 07.0 	&	1.720E+5	&	2.866E+4	&	0.9189	&	0.0135	&	0.00185	&	  18.883	&	   3.074\\
Hen 2-15	&	08 53 30.91 	&	-40 03 42.3 	&	2.074E+5	&	1.167E+5	&	0.7399	&	0.1666	&	0.00152	&	  20.618	&	   9.767\\
NGC 2867	&	09 21 25.38 	&	-58 18 40.9 	&	1.396E+6	&	2.150E+5	&	0.5991	&	0.0750	&	0.01265	&	 145.869	&	  97.873\\
IC 2501		&	09 38 47.146 	&	-60 05 30.52 	&	5.510E+6	&	6.918E+5	&	0.7746	&	0.0284	&	0.06262	&	 594.869	&	 249.031\\
NGC 3132	&	10 07 01.764 	&	-40 26 11.12 	&	2.046E+5	&	8.553E+4	&	0.9436	&	0.0232	&	0.00226	&	  22.607	&	   2.595\\
Hen 2-47	&	10 23 09.14 	&	-60 32 42.3 	&	6.968E+6	&	5.956E+5	&	0.5406	&	0.0459	&	0.07522	&	 739.082	&	 547.105\\
NGC 3242	&	10 24 46.107 	&	-18 38 32.64 	&	5.172E+5	&	5.306E+4	&	0.6945	&	0.0250	&	0.00486	&	  54.735	&	  29.656\\
IC 2621		&	11 00 19.99 	&	-65 14 57.8 	&	2.760E+7	&	1.244E+6	&	0.7107	&	0.0143	&	0.21028	&	2503.267	&	1296.595\\
NGC 3587	&	11 14 47.701 	&	+55 01 08.72 	&	1.550E+3	&	8.467E+4	&	0.0000	&		&	0.00002	&	   0.166	&	   0.174\\
NGC 3918	&	11 50 17.77 	&	-57 10 56.4 	&	3.916E+6	&	4.138E+5	&	0.7457	&	0.0292	&	0.03210	&	 394.906	&	 183.515\\
NGC 4361	&	12 24 30.76 	&	-18 47 05.4 	&	1.884E+6	&	4.679E+5	&	0.9974	&	0.0005	&	0.00835	&	 163.881	&	   0.876\\
IC 3568		&	12 33 06.871 	&	+82 33 48.95 	&	6.186E+5	&	1.398E+5	&	0.8415	&	0.0380	&	0.00649	&	  67.305	&	  20.559\\
NGC 5307	&	13 51 03.322 	&	-51 12 20.77 	&	3.122E+5	&	1.570E+5	&	0.5694	&	0.2821	&	0.00251	&	  31.779	&	  22.478\\
NGC 5315	&	13 53 57.00 	&	-66 30 50.7 	&	8.513E+6	&	7.811E+5	&	0.8083	&	0.0172	&	0.10529	&	 919.434	&	 333.545\\
Hen 2-113	&	14 59 53.476 	&	-54 18 07.42 	&	5.718E+7	&	2.292E+7	&	0.0000	&		&	0.61588	&	4690.332	&	4910.109\\
Hen 2-142	&	15 59 57.63 	&	-55 55 33.0 	&	9.975E+6	&	1.195E+6	&	0.7862	&	0.0258	&	0.10745	&	1039.136	&	 415.443\\
IC 4593		&	16 11 44.544 	&	+12 04 17.06 	&	1.578E+6	&	5.534E+5	&	0.9192	&	0.0255	&	0.02040	&	 179.739	&	  29.183\\
NGC 6072	&	16 12 58.40 	&	-36 13 47.0 	&	6.337E+4	&	3.877E+4	&	0.8552	&	0.0945	&	0.00059	&	   6.701	&	   1.884\\
NGC 6153	&	16 31 30.626 	&	-40 15 12.31 	&	1.603E+6	&	3.786E+5	&	0.7782	&	0.0563	&	0.01541	&	 169.767	&	  70.093\\
NGC 6210	&	16 44 29.491 	&	+23 47 59.68 	&	9.176E+5	&	6.085E+5	&	0.0000	&		&	0.01057	&	 102.019	&	 106.778\\
IC 4634		&	17 01 33.57 	&	-21 49 33.3 	&	1.171E+6	&	1.229E+5	&	0.7378	&	0.0286	&	0.01331	&	 129.607	&	  61.816\\
NGC 6302	&	17 13 44.339 	&	-37 06 10.95 	&	9.779E+6	&	2.545E+6	&	0.9678	&	0.0073	&	0.05513	&	 884.954	&	  58.724\\
NGC 6309	&	17 14 04.299 	&	-12 54 35.74 	&	3.615E+5	&	8.531E+4	&	0.4552	&	0.1719	&	0.00339	&	  38.289	&	  31.772\\
H 1-13		&	17 28 27.503 	&	-35 07 31.58 	&	3.514E+6	&	1.464E+6	&	0.5394	&	0.2006	&	0.03785	&	 378.898	&	 281.174\\
NGC 6369	&	17 29 20.45 	&	-23 45 34.8 	&	2.369E+6	&	5.004E+5	&	0.5832	&	0.1157	&	0.01942	&	 240.382	&	 166.032\\
Hb 4		&	17 41 52.80 	&	-24 42 08.7 	&	3.319E+6	&	4.091E+5	&	0.7387	&	0.0346	&	0.01942	&	 353.915	&	 168.292\\
Hb 5		&	17 47 56.20 	&	-29 59 39.6 	&	2.668E+7	&	6.677E+6	&	0.9933	&	0.0007	&	0.03396	&	2404.431	&	  33.666\\
NGC 6445	&	17 49 15.21 	&	-20 00 34.5 	&	1.858E+6	&	6.527E+5	&	0.9934	&	0.0022	&	0.01665	&	 193.196	&	   2.675\\
NGC 6543	&	17 58 33.423 	&	+66 37 59.52 	&	1.112E+6	&	7.615E+4	&	0.5645	&	0.0379	&	0.01630	&	 130.998	&	  93.416\\
NGC 6537	&	18 05 13.104 	&	-19 50 34.88 	&	1.126E+7	&	9.717E+5	&	0.8038	&	0.0164	&	0.06789	&	1030.472	&	 381.725\\
M 1-40		&	18 08 25.989 	&	-22 16 52.93 	&	3.543E+6	&	2.567E+6	&	0.7090	&	0.2182	&	0.02936	&	 358.767	&	 186.746\\
M 1-41		&	18 09 30.10 	&	-24 12 26.0 	&	3.087E+6	&	2.685E+6	&	0.9940	&	0.0042	&	0.03042	&	 326.550	&	   4.080\\
NGC 6572	&	18 12 06.365 	&	+06 51 13.01 	&	1.532E+7	&	1.172E+6	&	0.8508	&	0.0111	&	0.16286	&	1547.986	&	 447.504\\
SwSt 1		&	18 16 12.268 	&	-30 52 08.01 	&	6.488E+7	&	6.079E+6	&	0.9599	&	0.0029	&	0.69886	&	5129.055	&	 422.482\\
NGC 6578	&	18 16 16.517 	&	-20 27 02.67 	&	9.625E+5	&	8.108E+5	&	0.5474	&	0.5162	&	0.01037	&	 105.203	&	  77.117\\
Cn 3-1		&	18 17 34.113 	&	+10 09 03.46 	&	7.673E+6	&	2.850E+6	&	0.8544	&	0.0512	&	0.03022	&	 639.986	&	 180.845\\
NGC 6629	&	18 25 42.458 	&	-23 12 10.23 	&	1.613E+6	&	7.764E+5	&	0.7826	&	0.1069	&	0.02117	&	 184.324	&	  74.756\\
M 2-43		&	18 26 40.05 	&	-02 42 57.3 	&	1.146E+8	&	1.381E+7	&	0.5506	&	0.0539	&	1.23460	&	7232.283	&	5277.104\\
M 1-51		&	18 33 29.05 	&	-11 07 26.5 	&	3.911E+6	&	1.599E+6	&	0.7893	&	0.0858	&	0.04212	&	 420.741	&	 166.031\\
M 1-59		&	18 43 20.20 	&	-09 04 49.1 	&	6.214E+6	&	1.119E+6	&	0.8287	&	0.0263	&	0.06693	&	 660.372	&	 216.476\\
M 1-61		&	18 45 55.12 	&	-14 27 37.9 	&	3.085E+7	&	2.752E+6	&	0.3319	&	0.0728	&	0.33230	&	2883.860	&	2686.219\\
Hu 2-1		&	18 49 47.567 	&	+20 50 39.45 	&	1.942E+7	&	1.565E+6	&	0.9028	&	0.0051	&	0.22686	&	1948.460	&	 377.028\\
NGC 6720	&	18 53 35.079 	&	+33 01 45.03 	&	1.438E+5	&	4.327E+4	&	0.8529	&	0.0474	&	0.00138	&	  15.336	&	   4.375\\
K 3-17		&	18 56 18.164 	&	+07 07 25.88 	&	3.817E+6	&	9.507E+5	&	0.8458	&	0.0379	&	0.04111	&	 410.849	&	 122.363\\
M 1-66		&	18 58 26.24 	&	-01 03 45.6 	&	7.136E+6	&	2.707E+6	&	0.5779	&	0.1432	&	0.07687	&	 754.659	&	 526.094\\
NGC 6741	&	19 02 37.10 	&	-00 26 56.7 	&	1.034E+7	&	2.005E+6	&	0.8794	&	0.0179	&	0.08388	&	1013.822	&	 240.583\\
Abell 53	&	19 06 45.910 	&	+06 23 52.47 	&	1.868E+7	&	3.379E+6	&	0.9986	&	0.0001	&	0.15997	&	1788.741	&	   5.234\\
Hen 2-430	&	19 14 04.13 	&	+17 31 33.1 	&	1.581E+7	&	7.609E+5	&	0.9084	&	0.0024	&	0.17035	&	1597.668	&	 292.355\\
NGC 6781	&	19 18 28.085 	&	+06 32 19.29 	&	1.358E+4	&	4.336E+5	&	0.0000	&		&	0.00015	&	   1.502	&	   1.572\\
NGC 6790	&	19 22 56.966 	&	+01 30 46.46 	&	6.330E+7	&	3.213E+6	&	0.8698	&	0.0055	&	0.50259	&	5056.511	&	1288.885\\
Vy 2-2		&	19 24 22.223 	&	+09 53 56.29 	&	6.476E+8	&	8.626E+7	&	0.9802	&	0.0021	&	6.97563	&	10190.46	&	 419.323\\
K 3-35		&	19 27 44.02 	&	+21 30 03.4 	&	2.712E+7	&	4.541E+6	&	0.9629	&	0.0036	&	0.29215	&	2584.072	&	 196.957\\
PB 10		&	19 28 14.487 	&	+12 19 37.11 	&	1.683E+6	&	5.816E+5	&	0.8975	&	0.0336	&	0.01812	&	 183.209	&	  37.305\\
NGC 6803	&	19 31 16.47 	&	+10 03 21.7 	&	7.843E+6	&	1.006E+6	&	0.8254	&	0.0170	&	0.09164	&	 840.595	&	 280.474\\
NGC 6804	&	19 31 35.14 	&	+09 13 31.4 	&	3.116E+6	&	2.165E+6	&	0.9954	&	0.0029	&	0.02929	&	 325.928	&	   3.138\\
BD+30 3639	&	19 34 45.233 	&	+30 30 58.94 	&	2.083E+7	&	1.113E+6	&	0.6156	&	0.0230	&	0.22435	&	2049.857	&	1332.475\\
M 1-71		&	19 36 26.92 	&	+19 42 24.1 	&	1.768E+7	&	6.701E+5	&	0.7992	&	0.0066	&	0.19561	&	1776.664	&	 671.951\\
M 1-73		&	19 41 09.29 	&	+14 56 58.8 	&	5.391E+5	&	9.091E+5	&	0.1400	&		&	0.00581	&	  59.059	&	  60.604\\
NGC 6818	&	19 43 58.022 	&	-14 09 13.44 	&	4.079E+5	&	5.174E+4	&	0.4568	&	0.0841	&	0.00324	&	  41.388	&	  34.278\\
NGC 6826	&	19 44 48.150 	&	+50 31 30.26 	&	7.624E+5	&	2.299E+5	&	0.6526	&	0.1269	&	0.00724	&	  80.840	&	  48.581\\
Hen 2-447	&	19 45 22.16 	&	+21 20 03.9 	&	1.886E+7	&	2.082E+6	&	0.6390	&	0.0360	&	0.20315	&	1875.197	&	1161.339\\
NGC 6842	&	19 55 02.1 	&	+29 17 22 	&	5.860E+5	&	1.929E+5	&	0.9915	&	0.0026	&	0.00310	&	  53.516	&	   0.953\\
Hen 1-4		&	19 59 18.014 	&	+31 54 39.14 	&	9.712E+6	&	1.249E+7	&	0.9981	&	0.0012	&	0.10743	&	1018.633	&	   4.074\\
K 3-52		&	20 03 11.44 	&	+30 32 34.1 	&	3.597E+7	&	3.695E+6	&	0.8536	&	0.0108	&	0.38751	&	3276.799	&	 930.736\\
M 1-75		&	20 04 44.1 	&	+31 27 28 	&	3.408E+5	&	3.187E+4	&	0.9782	&	0.0019	&	0.00367	&	  37.378	&	   1.690\\
K 3-55		&	20 06 56.250 	&	+32 16 36.80 	&	1.463E+6	&	4.549E+5	&	0.7405	&	0.0862	&	0.01576	&	 159.488	&	  75.407\\
NGC 6884	&	20 10 23.66 	&	+46 27 39.8 	&	3.662E+6	&	5.046E+5	&	0.6440	&	0.0526	&	0.03893	&	 393.269	&	 240.900\\
NGC 6881	&	20 10 52.45 	&	+37 24 42.4 	&	8.812E+6	&	2.081E+6	&	0.9300	&	0.0130	&	0.08176	&	 894.926	&	 126.555\\
NGC 6886	&	20 12 42.83 	&	+19 59 22.6 	&	4.327E+6	&	1.933E+6	&	0.8147	&	0.0783	&	0.03364	&	 430.087	&	 151.363\\
K 3-57		&	20 12 47.725 	&	+34 20 32.79 	&	7.122E+5	&	1.228E+6	&	0.0000	&		&	0.00767	&	  77.950	&	  81.586\\
Hen 2-459	&	20 13 57.89 	&	+29 33 55.9 	&	4.854E+7	&	7.524E+6	&	0.8936	&	0.0120	&	0.52288	&	4153.345	&	 876.039\\
NGC 6891	&	20 15 08.838 	&	+12 42 15.63 	&	2.839E+6	&	8.101E+5	&	0.9168	&	0.0213	&	0.03058	&	 307.163	&	  51.258\\
NGC 6894	&	20 16 23.965 	&	+30 33 53.17 	&	3.999E+5	&	2.284E+5	&	0.9844	&	0.0087	&	0.00431	&	  43.846	&	   1.423\\
IC 4997		&	20 20 08.742 	&	+16 43 53.71 	&	3.976E+7	&	9.787E+6	&	0.7952	&	0.0417	&	0.18786	&	3219.897	&	1238.914\\
M 3-35		&	20 21 03.77 	&	+32 29 24.0 	&	4.596E+7	&	3.791E+6	&	0.9272	&	0.0044	&	0.49506	&	3982.741	&	 585.173\\
NGC 6905	&	20 22 22.991 	&	+20 06 16.25 	&	2.252E+6	&	1.514E+6	&	0.9933	&	0.0041	&	0.01928	&	 231.096	&	   3.226\\
NGC 7008	&	21 00 32.7 	&	+54 32 39 	&	1.096E+5	&	2.465E+4	&	0.9138	&	0.0195	&	0.00118	&	  12.038	&	   2.078\\
NGC 7009	&	21 04 10.877 	&	-11 21 48.26 	&	8.012E+5	&	5.440E+4	&	0.3776	&	0.0552	&	0.00911	&	  88.827	&	  79.713\\
NGC 7026	&	21 06 18.237 	&	+47 51 07.15 	&	9.590E+5	&	7.862E+4	&	0.8750	&	0.0075	&	0.01204	&	 108.877	&	  26.709\\
NGC 7027	&	21 07 01.8 	&	+42 14 10 	&	5.450E+7	&	9.574E+6	&	0.6965	&	0.0630	&	0.45157	&	4505.795	&	2428.366\\
NGC 7048	&	21 14 15.25 	&	+46 17 16.1 	&	1.885E+5	&	1.909E+5	&	0.9820	&	0.0174	&	0.00153	&	  19.248	&	   0.719\\
K 3-62		&	21 31 50.18 	&	+52 33 51.6 	&	9.779E+6	&	1.930E+6	&	0.7430	&	0.0418	&	0.10534	&	1019.792	&	 478.198\\
IC 5117		&	21 32 30.97 	&	+44 35 47.5 	&	6.225E+7	&	2.563E+6	&	0.7291	&	0.0092	&	0.56422	&	5001.907	&	2452.423\\
Hu 1-2		&	21 33 08.328 	&	+39 38 09.72 	&	3.021E+6	&	3.404E+6	&	0.0000	&		&	0.03529	&	 332.913	&	 348.447\\
Bl 2-1		&	22 20 16.63 	&	+58 14 16.6 	&	1.637E+7	&	2.118E+6	&	0.4362	&	0.0782	&	0.17631	&	1648.746	&	1397.387\\
IC 5217		&	22 23 55.73 	&	+50 58 00.5 	&	1.504E+6	&	9.728E+5	&	0.7386	&	0.1812	&	0.01363	&	 157.056	&	  74.700\\
Me 2-2		&	22 31 43.683 	&	+47 48 03.91 	&	2.123E+7	&	1.951E+6	&	0.7895	&	0.0127	&	0.21177	&	2061.190	&	 812.550\\
NGC 7354	&	22 40 19.83 	&	+61 17 08.7 	&	7.675E+5	&	1.342E+5	&	0.7503	&	0.0446	&	0.00650	&	  79.049	&	  36.155\\
NGC 7662	&	23 25 53.6 	&	+42 32 06 	&	8.681E+5	&	1.240E+5	&	0.6821	&	0.0508	&	0.00689	&	  87.918	&	  49.206\\
\hline
\end{longtable}

\section{Spectra of individual PNe}

   \begin{figure*}
   \resizebox{0.92\hsize}{!}
            {\includegraphics{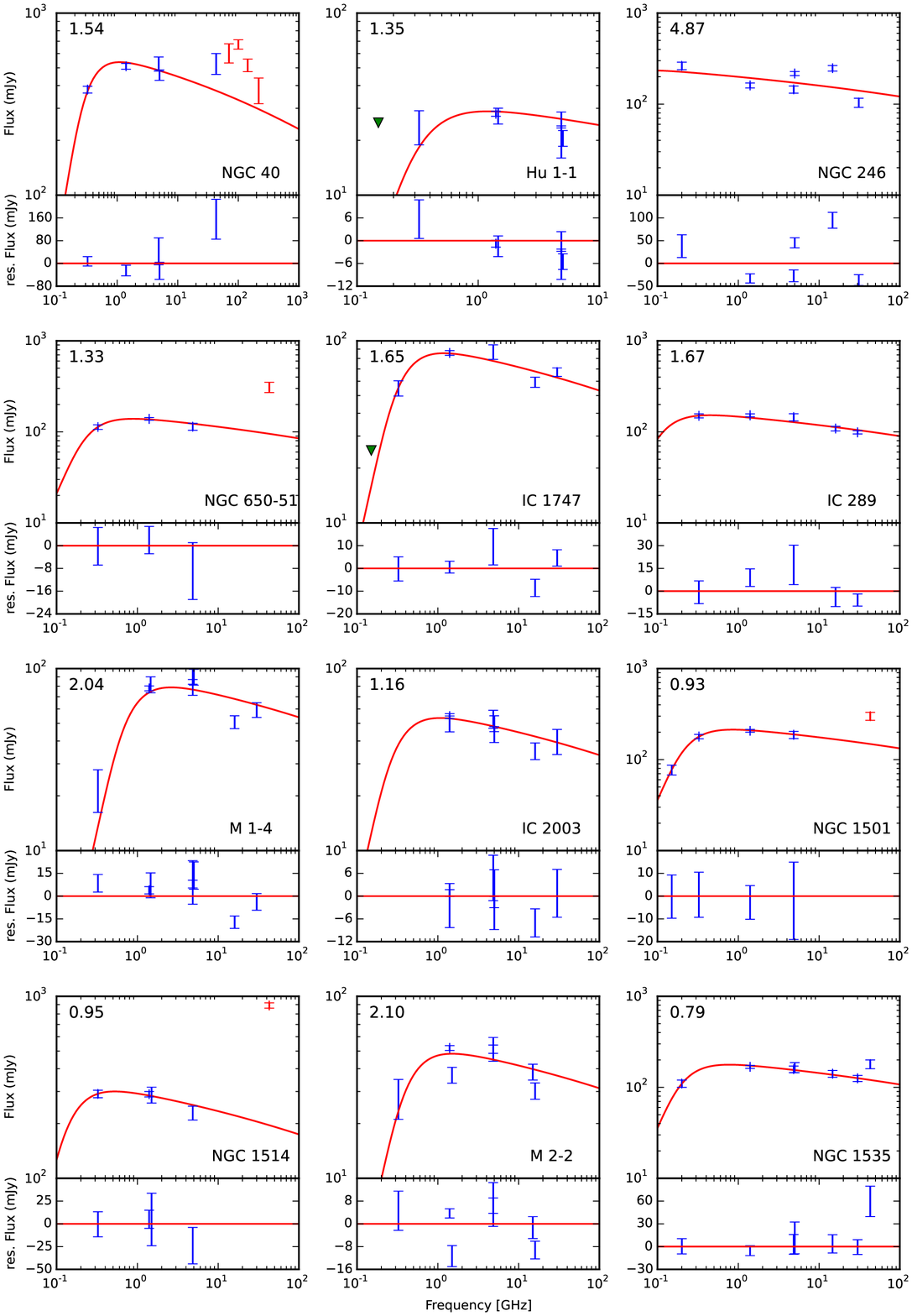}}
      \caption{Radio spectra and best model fits for PNe radio spectra. Triangles set an upper limit for the TGSS survey, if it is lower than the next data point. Red errorbars mark data not used in the fit.
              }
         \label{spec1}
   \end{figure*}

   \begin{figure*}
   \resizebox{0.92\hsize}{!}
            {\includegraphics{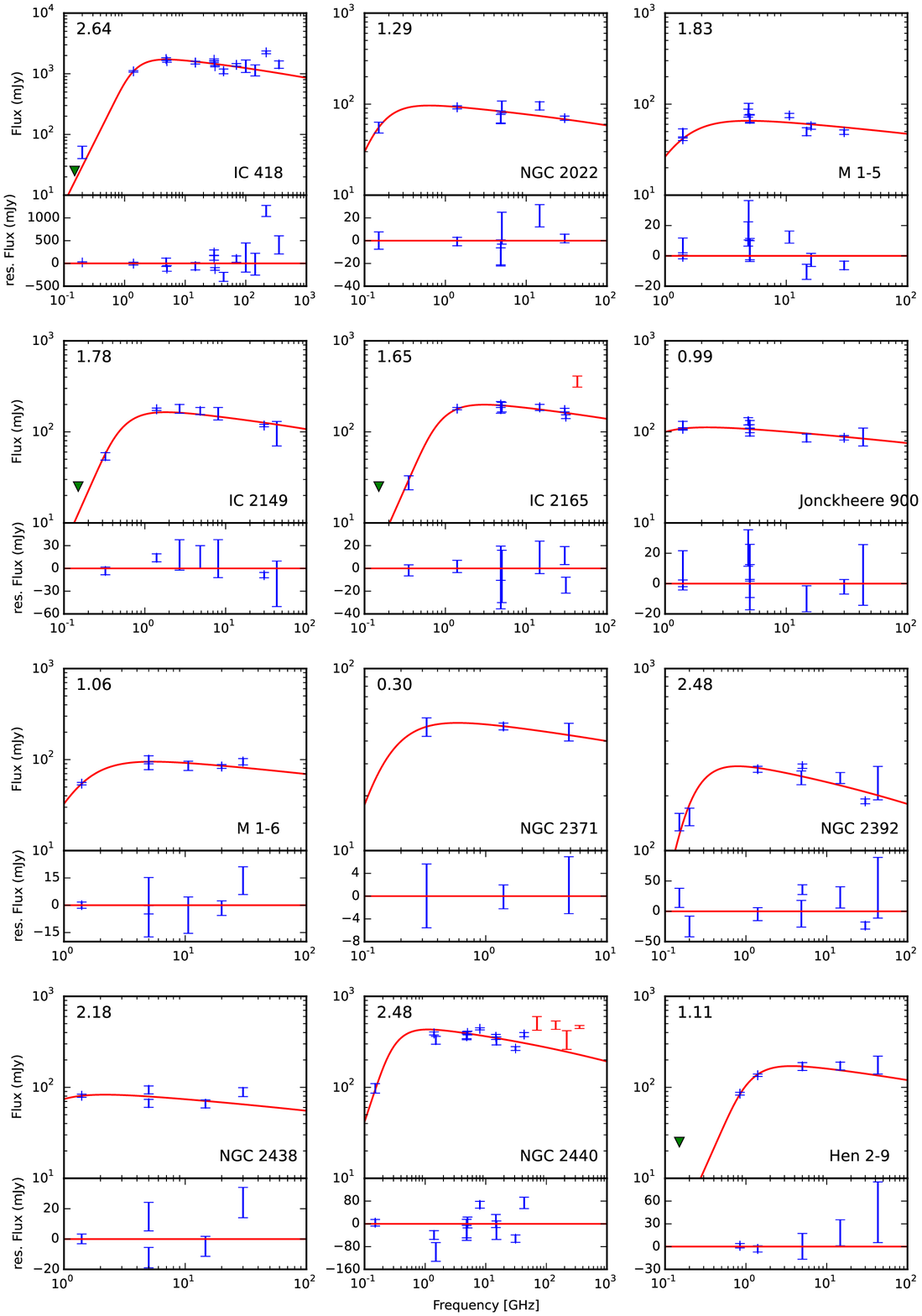}}
      \caption{Radio spectra and best model fits for PNe radio spectra. Triangles set an upper limit for the TGSS survey, if it is lower than the next data point. Red errorbars mark data not used in the fit.
              }
         \label{spec2}
   \end{figure*}

   \begin{figure*}
   \resizebox{0.92\hsize}{!}
            {\includegraphics{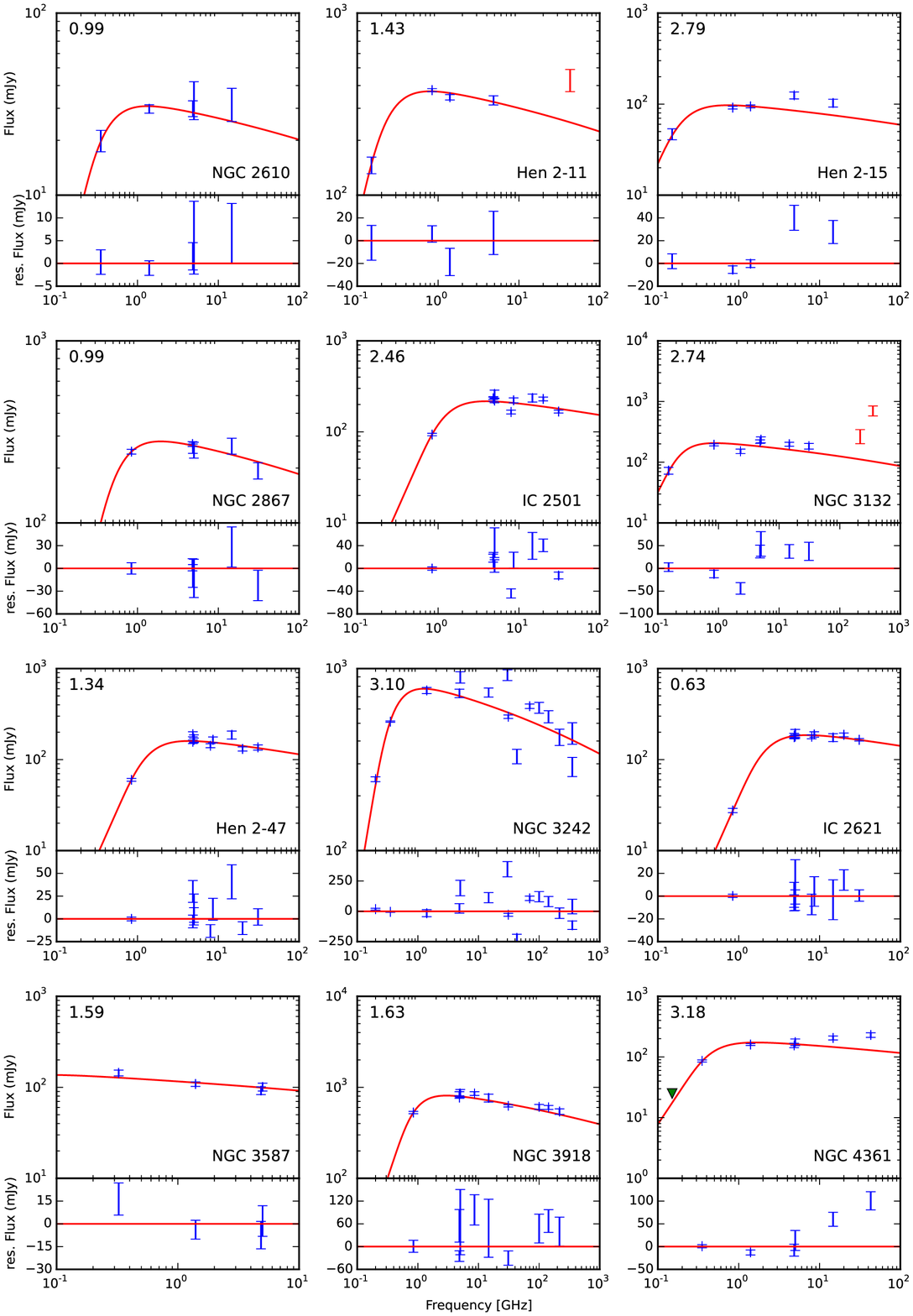}}
      \caption{Radio spectra and best model fits for PNe radio spectra. Triangles set an upper limit for the TGSS survey, if it is lower than the next data point. Red errorbars mark data not used in the fit.
              }
         \label{spec3}
   \end{figure*}

   \begin{figure*}
   \resizebox{0.92\hsize}{!}
            {\includegraphics{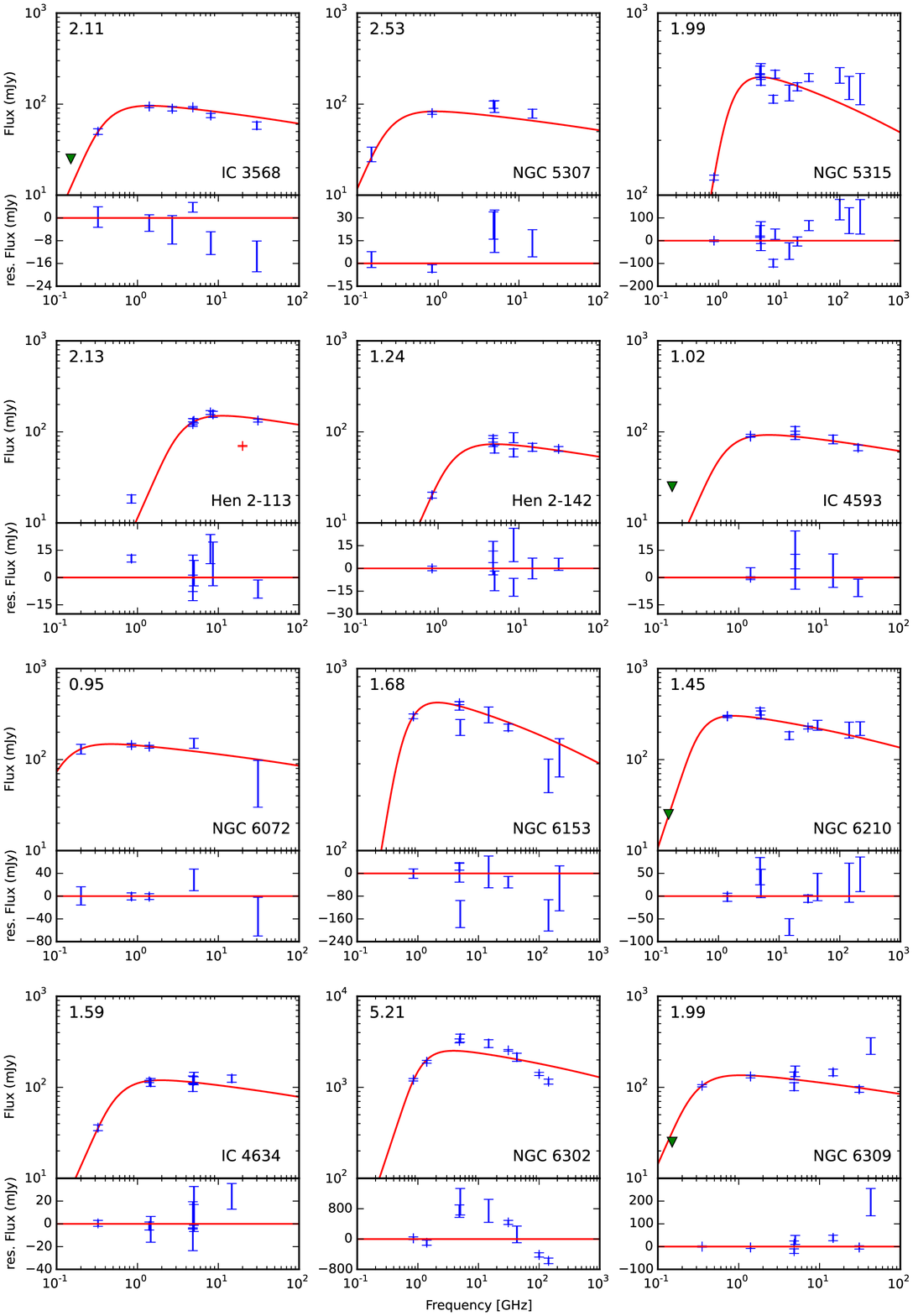}}
      \caption{Radio spectra and best model fits for PNe radio spectra. Triangles set an upper limit for the TGSS survey, if it is lower than the next data point. Red errorbars mark data not used in the fit.
              }
         \label{spec4}
   \end{figure*}

   \begin{figure*}
   \resizebox{0.92\hsize}{!}
            {\includegraphics{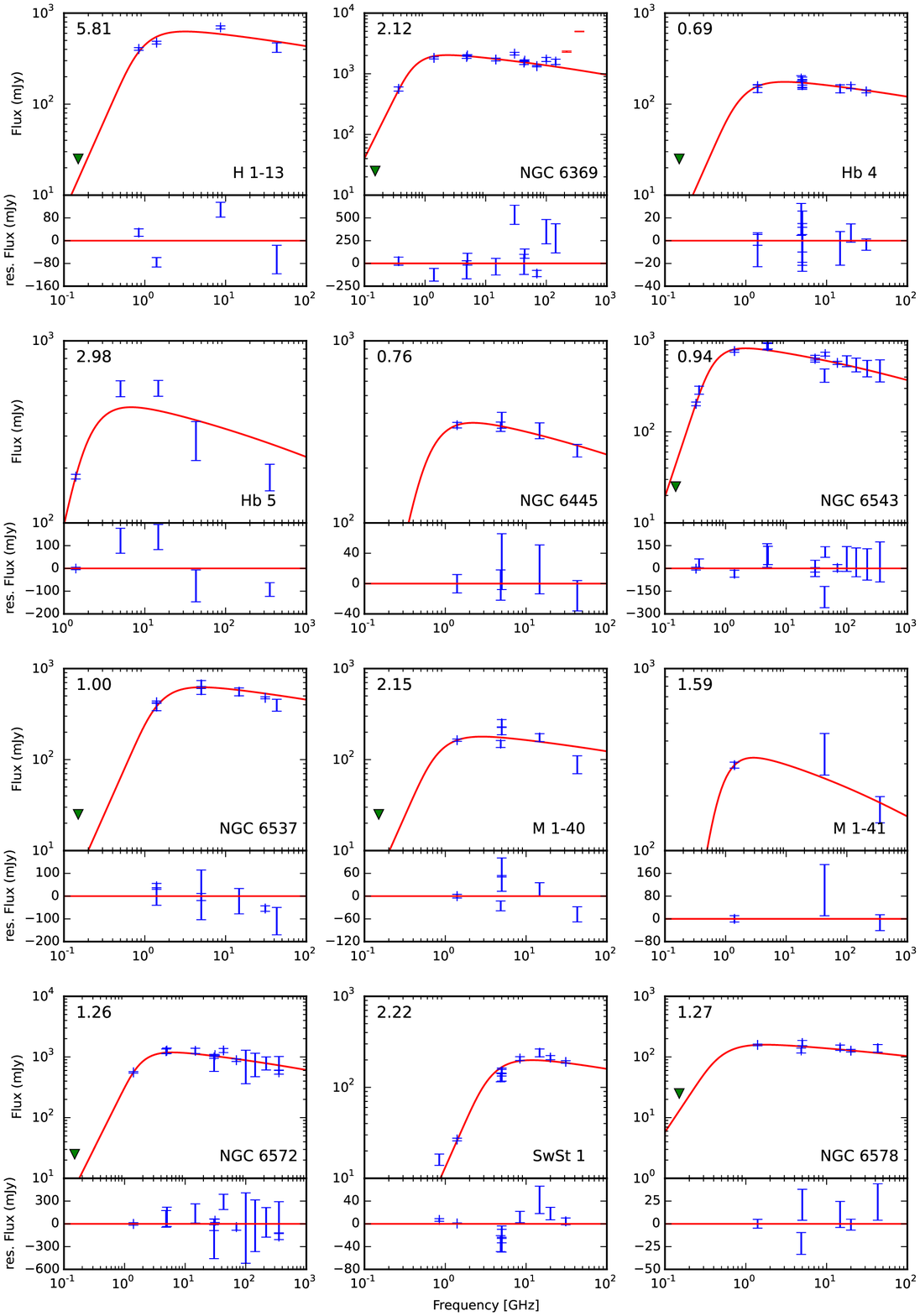}}
      \caption{Radio spectra and best model fits for PNe radio spectra. Triangles set an upper limit for the TGSS survey, if it is lower than the next data point. Red errorbars mark data not used in the fit.
              }
         \label{spec5}
   \end{figure*}

   \begin{figure*}
   \resizebox{0.92\hsize}{!}
            {\includegraphics{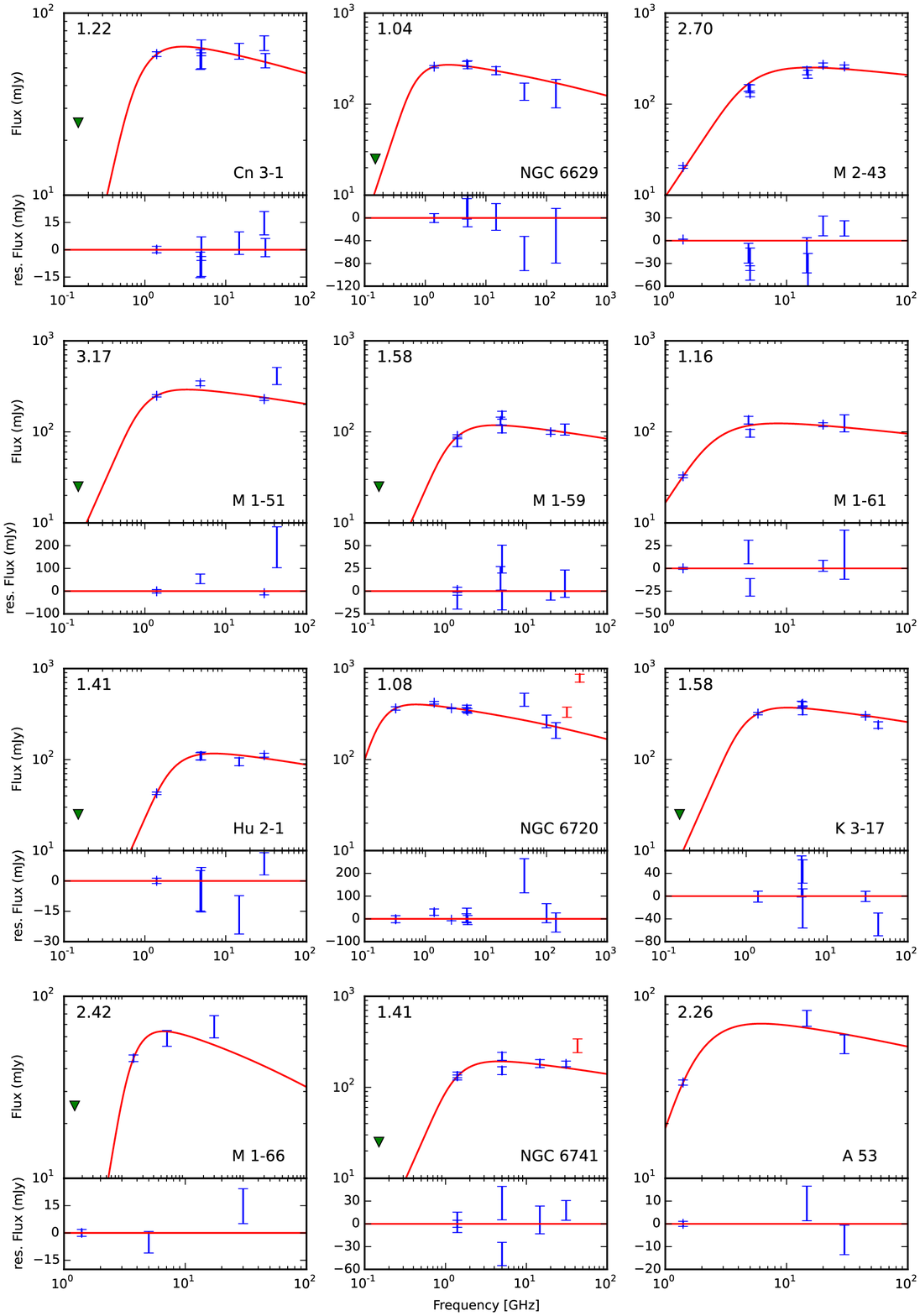}}
      \caption{Radio spectra and best model fits for PNe radio spectra. Triangles set an upper limit for the TGSS survey, if it is lower than the next data point. Red errorbars mark data not used in the fit.
              }
         \label{spec6}
   \end{figure*}

\begin{figure*}
   \resizebox{0.92\hsize}{!}
            {\includegraphics{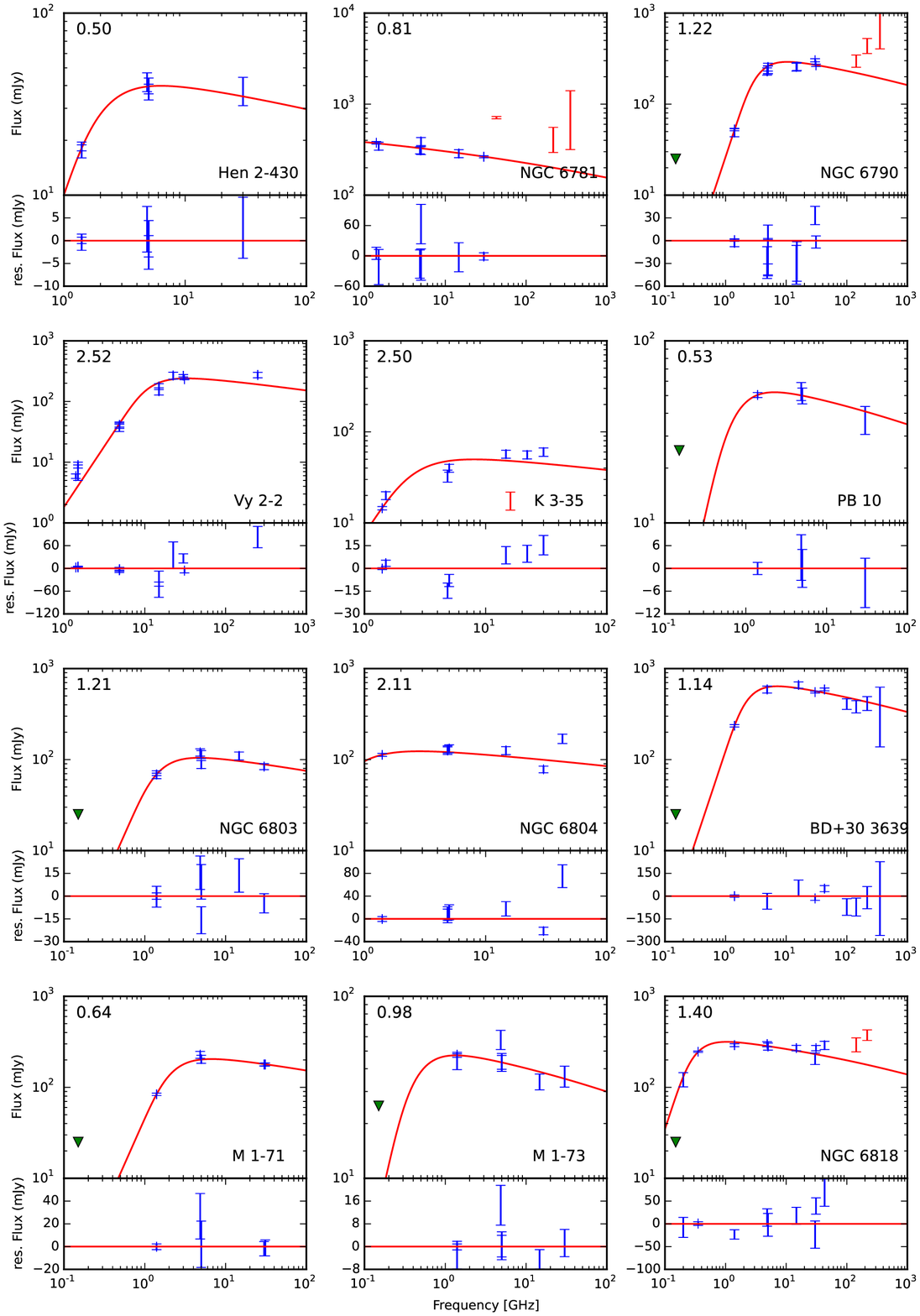}}
      \caption{Radio spectra and best model fits for PNe radio spectra. Triangles set an upper limit for the TGSS survey, if it is lower than the next data point. Red errorbars mark data not used in the fit.
              }
         \label{spec7}
   \end{figure*}

   \begin{figure*}
   \resizebox{0.92\hsize}{!}
            {\includegraphics{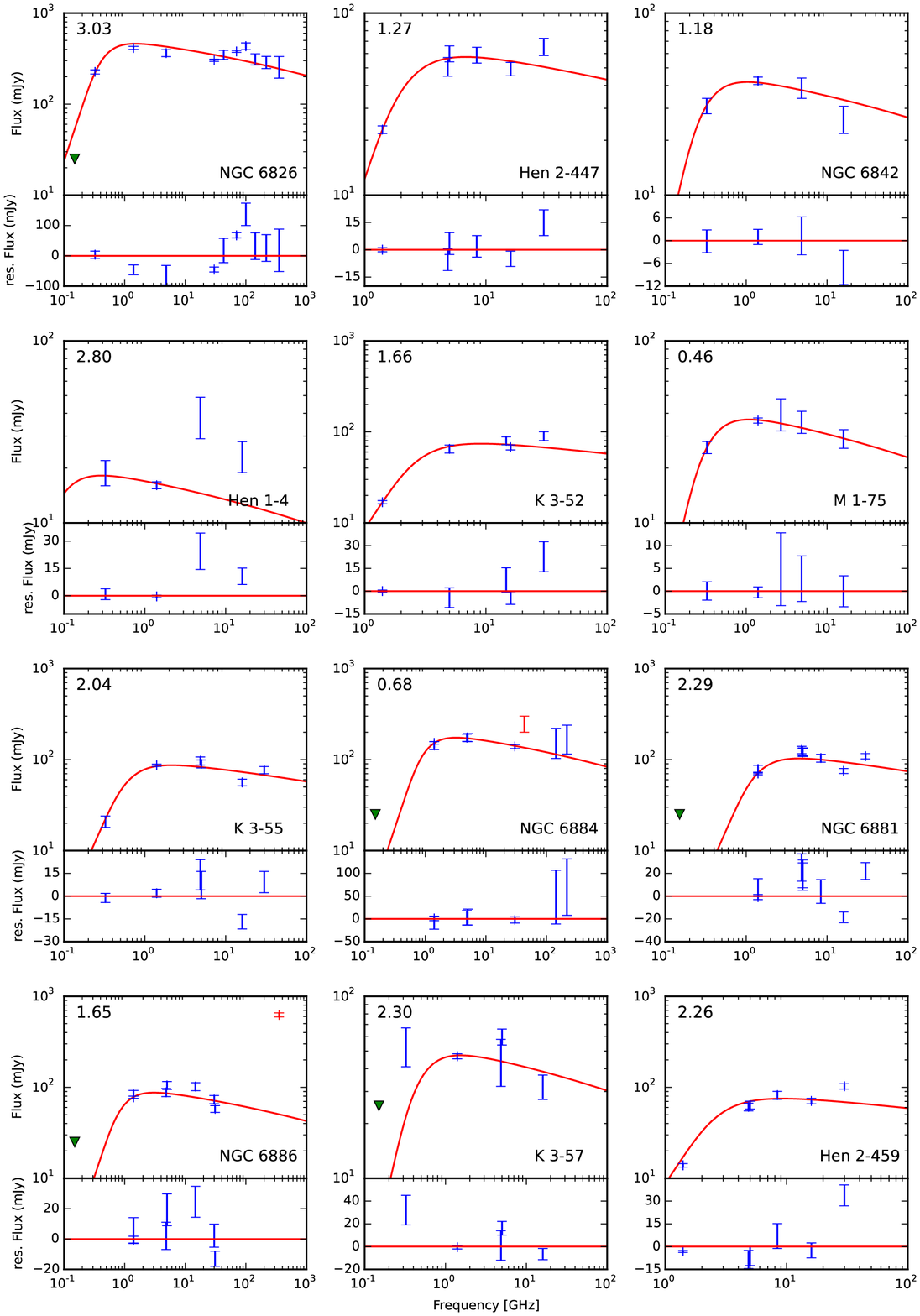}}
      \caption{Radio spectra and best model fits for PNe radio spectra. Triangles set an upper limit for the TGSS survey, if it is lower than the next data point. Red errorbars mark data not used in the fit.
              }
         \label{spec8}
   \end{figure*}

\begin{figure*}
   \resizebox{0.92\hsize}{!}
            {\includegraphics{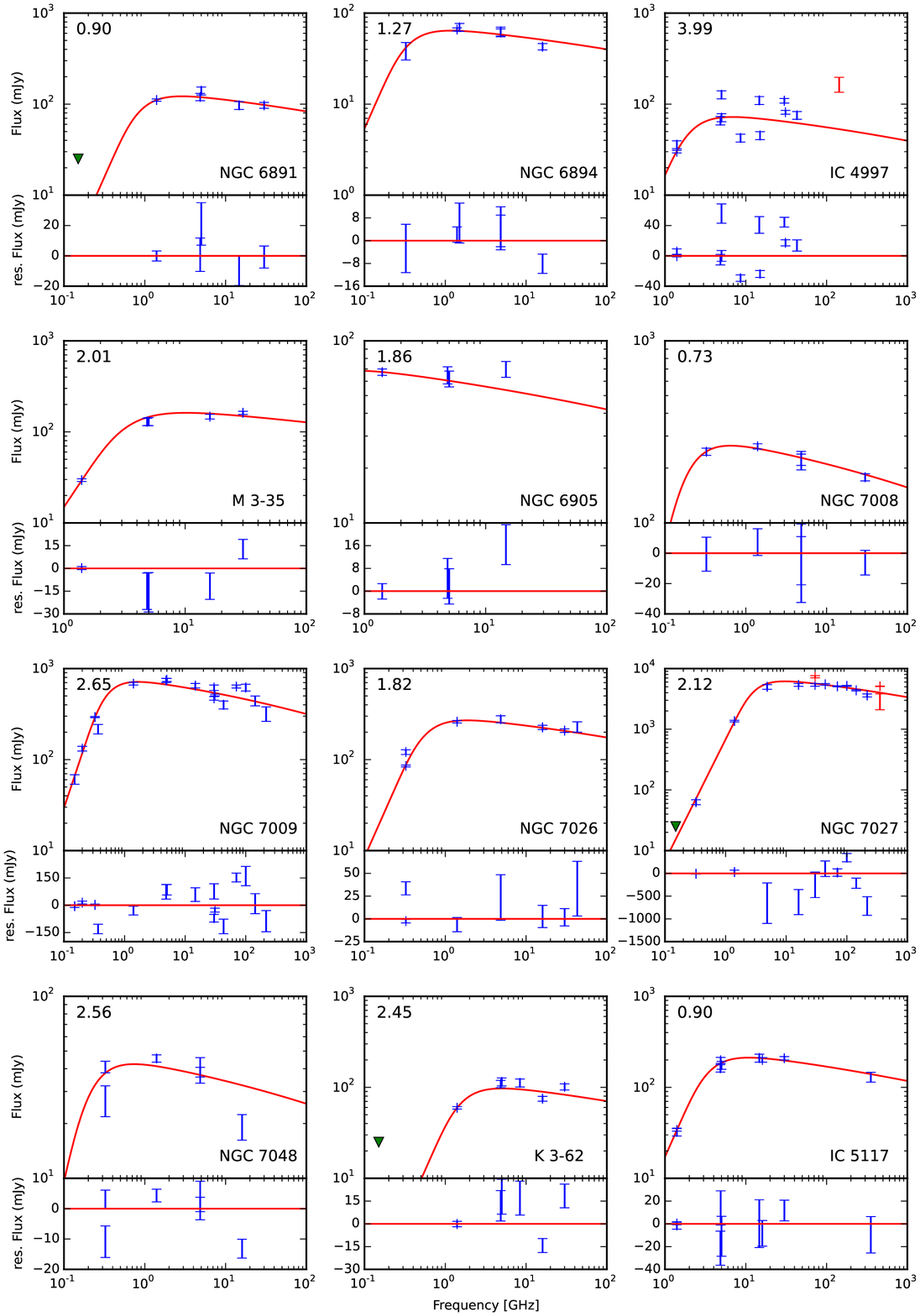}}
      \caption{Radio spectra and best model fits for PNe radio spectra. Triangles set an upper limit for the TGSS survey, if it is lower than the next data point. Red errorbars mark data not used in the fit.
              }
         \label{spec9}
   \end{figure*}

   \begin{figure*}
   \resizebox{0.92\hsize}{!}
            {\includegraphics{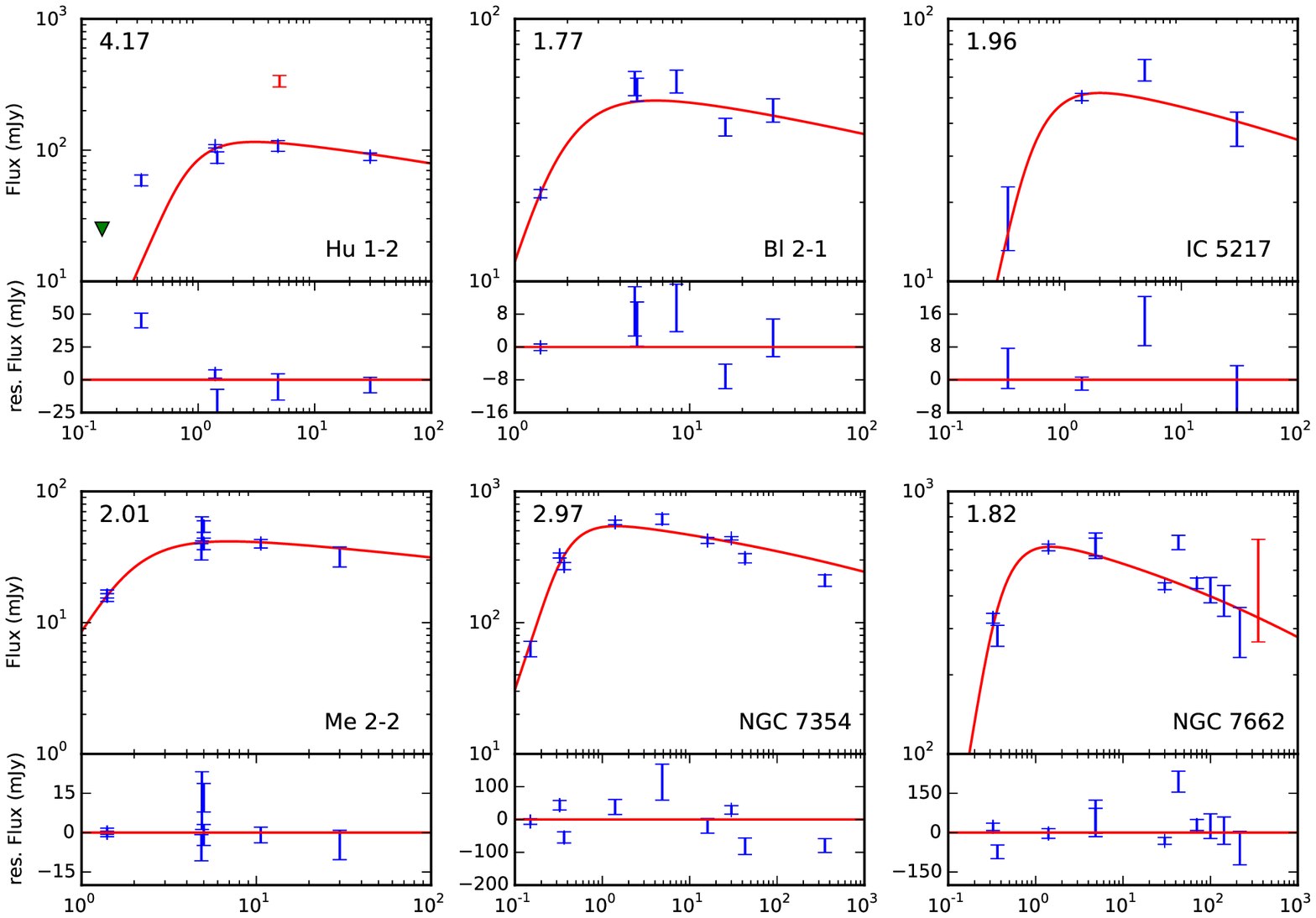}}
      \caption{Radio spectra and best model fits for PNe radio spectra. Triangles set an upper limit for the TGSS survey, if it is lower than the next data point. Red errorbars mark data not used in the fit.
              }
         \label{spec10}
   \end{figure*}

\bsp	
\label{lastpage}
\end{document}